\documentclass{emulateapj}

\usepackage{graphicx}
\usepackage{amsmath}
\usepackage{natbib}
\usepackage{amssymb}

\usepackage[breaklinks,colorlinks,urlcolor=blue,citecolor=blue,linkcolor=blue]{hyperref}

\usepackage{graphicx}  
\usepackage{dcolumn}   
\usepackage{bm}        
\usepackage{amsfonts,amsmath,amssymb,mathrsfs}
\usepackage{color}

\usepackage{hyperref}
\hypersetup{
  colorlinks=true,        
  linkcolor=blue,         
  citecolor=cyan,         
}
\usepackage{natbib,times}
\definecolor{orange}{rgb}{1,0.5,0}

\newcommand{\ttt}[1]{\texttt{\small #1}}
\newcommand{\tbf}[1]{#1}
\newcommand{\cf}{cf.~}
\newcommand{\ie}{i.e.,~}
\newcommand{\eg}{e.g.,~}

\begin{document}

\title{Electromagnetic emission
  from blitzars and its impact on non-repeating fast radio bursts}





\author{Elias R. Most\altaffilmark{1}, Antonios
  Nathanail\altaffilmark{1}, and Luciano Rezzolla\altaffilmark{1,2}}
\altaffiltext{1}{Institut f\"ur Theoretische Physik, Goethe Universit\"at Frankfurt,
Max-von-Laue-Str. 1, 60438 Frankfurt am Main, Germany}
\altaffiltext{2}{Frankfurt Institute for Advanced Studies, Ruth-Moufang-Str. 1, 60438
Frankfurt am Main, Germany}





\begin{abstract}
  \tbf{
It has been suggested that a non-repeating fast radio burst (FRB)
represents the final signal of a magnetized neutron star collapsing
to a black hole. In this model, a supramassive neutron star supported by
rapid rotation, will collapse to a black hole several thousand to million
years after its birth as a result of spin down. The collapse violently
snaps the magnetic-field lines anchored on the stellar surface, thus
producing an electromagnetic pulse that will propagate outwards and
accelerate electrons producing a massive radio burst, \ie a
``blitzar''. We present a systematic study of the gravitational collapse
of rotating and magnetised neutron stars with special attention to
far-field evolution at late times after the collapse. By considering a
series of neutron stars with rotation ranging from zero to millisecond
periods and different magnetic-field strengths, we show that the blitzar
emission is very robust and always characterised by a series
sub-millisecond pulses decaying exponentially in amplitude. The
luminosity and energy released when the magnetosphere is destroyed are
well reproduced by a simple expression in terms of the stellar magnetic
field and radius. Finally, we assess the occurrence of pair production
during a blitzar scenario, concluding that for typical magnetic-field
strengths of $10^{12}\,{\rm G}$ and spin frequencies of a few Hz, pair
production is suppressed. Overall, the very good match between the
results of the simulations and the luminosities normally observed for
FRBs lends credibility to the blitzar model as a simple and yet plausible
explanation for the phenomenology of non-repeating FRBs.
}
\end{abstract}

\keywords{
black hole physics -- MHD -- methods: numerical -- stars: neutron.
}



\section{Introduction}
\label{sec:intro}

The gravitational collapse of a magnetised and rotating neutron star can
lead to interesting and multimessenger emission, both in terms of
gravitational waves (GW) and in terms of electromagnetic (EM) radiation
in different bands. The detailed study of this combined emission can
provide important insight into a number of astrophysical observations
and, in particular, on a new type of astrophysical phenomena that is
collectively referred to as \textit{fast radio bursts (FRBs}, for a
review \cite{RaneLorimer2017}).

FRBs are bright, millisecond radio single pulses that do not normally
repeat and are not associated with a known pulsar or gamma-ray burst. The
accounted dispersion measurements suggest that they are extragalactic,
thus implying that their high radio luminosity is far larger than the
single pulses from known pulsars. Furthermore, evidence of high
magnetisation levels has been observed through Faraday rotation
measurements close to the source of a single FRB 110523 \citep{Masui2015} 
Because these transient radio sources are yet to be linked with confidence 
with a theoretical model, dozens of them exist in the literature, 
explaining either repeating FRBs such as \cite{Popov2007, Pen2015, 
Lyubarsky2014, Katz2016, Cordes2016, Lyutikov2016} or FRBs that have been 
detected only once and that represent the large majority \citep{Piro2012, 
Totani2013, Wang2016, Kashiyama2013, Mingarelli2015, Zhang2016, 
Liebling2016}.

The ``blitzar'' model of \citet{Falcke2013}, is particularly relevant for
our study as it involves the gravitational collapse of rotating and
magnetised neutron stars. More specifically, in this model, an isolated
and magnetised neutron star that was born massive enough, can be
supported against gravitational collapse by its rapid rotation. However,
the star is also continuously spinning down because of loss of kinetic
rotational energy in the form of EM dipolar emission; the spindown
continues up until the threshold to a dynamical instability to
gravitational collapse -- the neutral stability line -- is reached and
the star then collapses on a dynamical timescale \citep{Takami:2011,
  Weih2017}. During the collapse, the magnetic field that was previously
anchored on the surface of the star can either follow it as the surface
is trapped behind the event horizon, or propagate outwards in the form of
EM waves as the magnetic field lines are snapped. If the magnetic field
is initially dipolar, the structure of these EM waves will be
quadrupolar, with large magnetic blobs carrying away most of the EM
energy. Furthermore, the travelling large-scale magnetic shock that
propagate outwards can accelerate free electrons which will produce radio
signals dissipating the radiated energy \citep{Falcke2013}.

Together with EM radiation, the collapsing star will also produce GW
radiation even if perfectly axisymmetric and simply because its
quadrupole moment will have a nonzero time derivative. While we do not
intend to focus here on the GW emission from these events, since detailed
studies already exist [see, \eg \citet{Baiotti2007} for a comprehensive
  discussion], we will rather concentrate on how the EM
emission is produced during the collapse and how it propagates outwards
as EM waves following the destruction of the large-scale and ordered
magnetosphere. In particular, our aim is to follow the evolution of the
neutron star's magnetic field, both interior and exterior, during the 
collapse to a black hole and to quantify how the EM luminosity depends 
on the initial parameters of the neutron star, namely, its rotation rate 
and its magnetic-field strength.

We should recall that the collapse of a magnetised rotating neutron star
is an old problem that has been studied rather extensively in the past,
although only very recently in full general relativity. The first step
was considered by \cite{Wilson1975}, who simulated the collapse in an
ideal-magnetohydrodynamical (MHD) framework including the magnetic field
only in the interior of the star. While this work represented a first attempt to
simulate this process, the fact that the magnetic field was isolated to
the stellar interior had the consequence that no EM radiation was
produced in the process, since the magnetic field is dragged with the
collapsing matter and eventually hides behind a horizon. More recently,
\cite{Baumgarte02b2} investigated this scenario by considering the
perturbative dynamics of the EM fields on a dynamical spacetime given by
an Oppenheimer-Snyder collapse, that is, by the collapse of a nonrotating
dust cloud. Starting with a dipolar magnetic field, \cite{Baumgarte02b2}
especially analysed how the magnetosphere exterior to the star changes
during and mostly after the collapse of the star and the formation of the
black hole. Furthermore, they showed that the magnetic flux decays
exponentially in time after the collapse, following the quasi normal
modes of the newly formed black hole, leaving an unstructured magnetic
field in the vicinity of the black hole and outward propagating
EM waves in the far field. 

Much of this behaviour was later analysed and confirmed by
\citet{Dionysopoulou:2012pp}, who studied similar initial conditions of a
nonrotating neutron star, but where the spacetime was self-consistently
evolved via the solution of the Einstein equations and the EM fields
within a fully general-relativistic resistive-MHD framework. The
gravitational collapse of two magnetised and rotating neutron stars was
instead investigated by \citet{Lehner2011},
 who considered and contrasted two different and extreme
magnetospheric conditions: electrovacuum and force-free. In particular,
they showed that, in the force-free case, the magnetic flux threading the
event horizon completely vanishes within a millisecond from black-hole
formation, mostly because of reconnection. On the other hand, in the case
of the electrovacuum simulation. it was shown that the EM emission
depends weakly on rotation. Following the work of
\citet{Dionysopoulou:2012pp} in resistive MHD and extending it to the case
of rotating magnetised neutron stars in electrovacuum,
\citet{Nathanail2017} have recently shown that the initial charge of the
neutron star can be trapped behind the apparent horizon, thus forming a
charged black hole of Kerr-Newman type.

We here further explore the scenario investigated by
\citet{Nathanail2017} focussing mostly on the late time evolution of the
far field, where we can identify and analyse the millisecond-long EM
pulses produced during the gravitational collapse and the violent
disruption of the magnetosphere. Once produced in the vicinity of the
surface of the star and the apparent horizon that soon forms, these EM
pulses propagate outwards carrying enormous amounts of energy, with
magnitudes which are in good agreement with that associated with FRBs.
Furthermore, we here explore how the energetics of the emission depends
on the basic properties of the stars, \ie spin rate and magnetic-field
strength, and provide a simplified algebraic expressions that reproduces
well our results. 

The plan of the paper is the following one. In Sec. \ref{sec:nsaid} we
briefly review the numerical setup and how the initial data is computed.
In Sec. \ref{sec:pair-pro} we discuss 
the significance of our results 
due to pair creation during collapse.
The analysis of the numerical results follows in
Sec. \ref{sec:nres}. 
Finally, the discussion of the astrophysical impact
of the results and our conclusions are presented in
Sec. \ref{sec:conclusions}.

\begin{table*}
  \centering
  \begin{tabular}{|l||c|c|r|c|c|c|c|}
    \hline
    \hline
    \!\!\! &\!\!\! $M$  & $R_c$  &
    \!\!\! $f_{\rm spin}$  & $J/M^2$  &\!\!\! $B_{\rm pol}$ &\!\!\!
    $E_{_{\rm EM}}$ & $\boldsymbol{E}$ \\
    \!\!\! &\!\!\! $[M_\odot]$  & $[\mathrm{km}]$  &
    \!\!\! $[\mathrm{Hz}]$  & $-$  &\!\!\! $[10^{13}\,\mathrm{G}]$
    &\!\!\! $[{10^{43}\rm erg}]$ & $-$ \\
    \hline
    \!\!\!\ttt{F000.B13}  & \!\!\!\!\!\!2.096 & \!\!\!14.940 & \!\!\!0   & \!\!\!0.000 & \!\!\!6.09   & \!\!\!\!\!\!22.03     & zero \\
    \!\!\!\ttt{F001.B12}  & \!\!\!\!\!\!2.076 & \!\!\!15.546 & \!\!\!1   & \!\!\!0.0005 & \!\!\!0.217   & \!\!\!\!\!\!3.12{E}-2     & corotating \\
    \!\!\!\ttt{F010.B13}  & \!\!\!\!\!\!2.096 & \!\!\!14.941 & \!\!\!10  & \!\!\!0.005 & \!\!\!2.39   & \!\!\!\!\!\!3.50      & GR rot. sphere \\
    \!\!\!\ttt{F050.B13}  & \!\!\!\!\!\!2.097 & \!\!\!14.945 & \!\!\!50  & \!\!\!0.025 & \!\!\!2.29   & \!\!\!\!\!\!3.06      & corotating \\
    \!\!\!\ttt{F080.B13}  & \!\!\!\!\!\!2.095 & \!\!\!15.069 & \!\!\!80  & \!\!\!0.041 & \!\!\!2.28   & \!\!\!\!\!\!3.02      & corotating \\
    \!\!\!\ttt{F100.B13}  & \!\!\!\!\!\!2.084 & \!\!\!15.442 & \!\!\!100  & \!\!\!0.053 & \!\!\!2.22   & \!\!\!\!\!\!5.57    & GR rot. sphere \\    
    \!\!\!\ttt{F140.B13}  & \!\!\!\!\!\!2.095 & \!\!\!15.216 & \!\!\!140 & \!\!\!0.072 & \!\!\!2.27   & \!\!\!\!\!\!2.99      & corotating \\
    \!\!\!\ttt{F200.B13}  & \!\!\!\!\!\!2.092 & \!\!\!15.511 & \!\!\!200 & \!\!\!0.107 & \!\!\!2.26   & \!\!\!\!\!\!3.04      & corotating \\
    \!\!\!\ttt{F300.B13}  & \!\!\!\!\!\!2.081 & \!\!\!16.190 & \!\!\!300 & \!\!\!0.174 & \!\!\!2.23   & \!\!\!\!\!\!4.43      & GR rot. sphere \\
    \!\!\!\ttt{F400.B13}  & \!\!\!\!\!\!2.103 & \!\!\!16.404 & \!\!\!400 & \!\!\!0.234 & \!\!\!2.33   & \!\!\!\!\!\!2.85      & corotating \\
    \!\!\!\ttt{F450.B13}  & \!\!\!\!\!\!2.132 & \!\!\!16.069 & \!\!\!450 & \!\!\!0.251 & \!\!\!2.44   & \!\!\!\!\!\!3.61      & GR rot. sphere \\
    \!\!\!\ttt{F550.B13}  & \!\!\!\!\!\!2.144 & \!\!\!17.292 & \!\!\!550 & \!\!\!0.343 & \!\!\!2.50   & \!\!\!\!\!\!1.80      & corotating \\
    \!\!\!\ttt{F600.B13}  & \!\!\!\!\!\!2.179 & \!\!\!16.946 & \!\!\!600 & \!\!\!0.357 & \!\!\!2.63   & \!\!\!\!\!\!4.49      & GR rot. sphere \\
    \!\!\!\ttt{F800.B13}  & \!\!\!\!\!\!2.104 & \!\!\! 9.909 & \!\!\!800 & \!\!\!0.314 & \!\!\!52.84  & \!\!\!\!\!\!0.87      & GR rot. sphere \\
    \hline                                                                                                                         
    \!\!\!\ttt{F500.B10}  & \!\!\!\!\!\!2.147 & \!\!\!16.210 & \!\!\!500 & \!\!\!0.281 & \!\!\!0.0025 & \!\!\!\!\!\!2.37{E}-6 & corotating \\
    \!\!\!\ttt{F500.B12}  & \!\!\!\!\!\!2.147 & \!\!\!16.210 & \!\!\!500 & \!\!\!0.281 & \!\!\!0.25   & \!\!\!\!\!\!2.37{E}-2 & corotating \\
    \!\!\!\ttt{F500.B14}  & \!\!\!\!\!\!2.147 & \!\!\!16.210 & \!\!\!500 & \!\!\!0.281 & \!\!\!25     & \!\!\!\!\!\!2.68{E}+2 & corotating \\
    \!\!\!\ttt{F500.B15}  & \!\!\!\!\!\!2.147 & \!\!\!16.210 & \!\!\!500 & \!\!\!0.281 & \!\!\!250    & \!\!\!\!\!\!4.13{E}+4 & corotating \\
   \hline
    \hline
  \end{tabular}
  \caption{Initial neutron-star models. Reported in the different columns
    are: the ADM mass $M$, the circumferential radius $R_c$, the spin
    frequency $f_{\rm spin}$, the dimensionless angular momentum with $J$
    the Komar angular momentum, the value of the magnetic field at the
    pole of the neutron star $B_{\rm pol}$, the radiated EM energy
    $E_{_{\rm EM}}$, and the type of initial electric-field
    configuration. Note that the value of the maximum magnetic field
    $B_c$ inside the neutron star, is a factor of $\sim 6$ larger in all
    models.}
  \label{tab:initial}
\end{table*}

\section{Numerical setup and initial data}
\label{sec:nsaid}

The simulations reported below have been performed using the
general-relativistic resistive-MHD code \ttt{WhiskyRMHD}
\citep{Dionysopoulou:2012pp, Dionysopoulou2015}. The code uses
high-resolution shock-capturing methods such as the Harten-Lax-van
Leer-Einfeldt (HLLE) approximate Riemann solver. Following
\citet{Nathanail2017}, we reconstruct our primitive variables at the cell
interfaces using the enhanced piecewise parabolic reconstruction (ePPM),
which does not reduce to first order at local maxima \citep{Colella2008,
  Reisswig2012b}. For the evolution of the spacetime, the
\ttt{WhiskyRMHD} code makes use of the Einstein Toolkit
framework~\citep{loeffler_2011_et}, which exploits the \ttt{McLachlan}
code for the space-time evolution and the \ttt{Carpet} driver for
fixed-box mesh refinement \citep{Schnetter-etal-03b}. An important point
in our resistive-MHD treatment is the calculation of the electric charge
$q$, which we compute at every timestep via the divergence of the
electric field, \ie $q= \nabla_i E^i$, as adopted in several other
works~\citep{Bucciantini2012a, Dionysopoulou:2012pp,
  Qian2016, Nathanail2017}.

Because the focus of our study is mainly the evolution of the magnetic
field in the exterior of the star and the luminosity produced during and
after the collapse, the use of a resistive-MHD framework is particularly
convenient. In particular, we can assume a negligibly small electrical
conductivity in the exterior of the star so that it can effectively
(although not exactly) reproduce an electrovacuum regime. At the same
time, we can use a very large value of the electrical conductivity in the
stellar interior so that we can reproduce the highly conducting matter.
However, connecting the two regimes of low and high conductivity across
the star and its exterior has the consequence that the set of
resistive-MHD equations becomes stiff and hence requires special time
stepping strategies. Following \citet{Palenzuela:2008sf}, we employ an
implicit-explicit Runge-Kutta time stepping (RKIMEX) algorithm
\citep{pareschi_2005_ier}, the details of our implementation can be
found in \citet{Dionysopoulou:2012pp} and \citet{Dionysopoulou2015}.
We use the same setup here as \citet{Nathanail2017} and choose a finest
resolution of $147\ \mathrm{m}$ with a total domain size of $1075\ \mathrm{km}$,
which in combination with the implicit time evolution scheme makes these
runs rather expensive.

The initial neutron-star models are computed using the \ttt{Magstar}
code \citep{Bocquet1995} of the \ttt{LORENE} library
(\url{www.lorene.obspm.fr}). In
particular, \ttt{Magstar} computes self-consistent uniformly rotating
neutron stars by solving the coupled system of the Einstein-Maxwell
equations.  In this way, and using a polytropic equation of state
\citep{Rezzolla_book:2013} $p=K \rho^\Gamma$ with $\Gamma=2$ and
$K=164.708$ we have computed a total of 17 initial stellar models, whose
properties are collected in Table \ref{tab:initial}. The evolutions have
been performed using a simple gamma-law $p = \rho \varepsilon \left(
\Gamma -1\right)$ to allow for shock heating. Note that each model is
characterised by a rotational frequency\footnote{Note that we choose an
  upper limit for the spinning frequency of $800\,{\rm Hz}$ because this
  is already higher than the fastest pulsar presently observed, \ie PSR
  J1748-2446ad, whose rotation period is $1.3\,{\rm ms}$
  \citep{Hessels2006}.} and by a magnetic-field strength\footnote{While
  a magnetic field of $10^{13}\,{\rm G}$ is about two orders of magnitude
  larger than what is normally expected in a blitzar.  As we will show in
  Sec. \ref{sec:em_emission}, the results follow a simple scaling
  relation with the magnetic-field strength, but using a large value
  reduces the computational costs.}; for instance, model
\ttt{F300.B13} refers to a magnetised neutron star with spin
frequency of $f=300\,{\rm Hz}$ and a dipolar magnetic \tbf{field} with a
value at the pole $B_{\rm pol}=10^{13}\,{\rm G}$.  Collectively, the
models presented in Table \ref{tab:initial} can be considered as
representative of the magnetic field strengths and of the rotational
frequencies to be expected by supramassive magnetised neutron stars just
before collapse.

Although the solution provided by \ttt{Magstar} includes
self-consistent electric and magnetic fields, a certain freedom remains
in the choice of an initial electric field that is consistent with our
electrovacuum prescription. In fact, while the electric field in the
neutron-star interior is always unambiguously given by the ideal-MHD
condition \ie $E^i = - \epsilon^{ijk} (v_{\rm c})_j B_k$, where $(v_{\rm
  c})_j$ is the corotation velocity and $\epsilon^{ijk}$ the totally
antisymmetric permutation symbol, the electric field outside the star
should be such that there are no charges outside, \ie the electric field
should be divergence free in the stellar exterior.  A similar ambiguity
was discussed by \citet{Nathanail2017}, who, after considering several
different options, found that the optimal initial electric field
minimising the exterior charge density is the one deriving from the
analytical solution of a rotating magnetised sphere in general relativity
\citep{Rezzolla2001, Rezzolla2001_err}.

Alternative approaches using a force-free description
of the magnetosphere have instead prescribed the electric field in terms
of the corotation velocity and of the magnetic field \citep{Lehner2011,
  Palenzuela2013}; this approach is appropriate inside the light cylinder
$r_{_L} := c/\Omega$ of such a magnetosphere.

Fortunately, the results of the simulations do not depend sensitively on
the choice made for the electric field and, as we show in Appendix
\ref{appen}, the variation of the EM luminosity with the different
prescriptions is minimal and the light curves overlap over the whole
duration of the intense EM emission. Hence we have opted for the most simple
and robust prescription of prescribing the electric field using the corotation
velocity as this does not require corrections for deviations from spherical 
symmetry in the case of the fast rotating models. Nonetheless, we have 
also performed some simulations with the rotating magnetised sphere prescription
for several models to give an error range to our calculations as detailed in
the Appendix \ref{appen}.

\section{Pair Production}
\label{sec:pair-pro}

Since we will be modelling the blitzar model in a purely electrovacuum
scenario, it is important to check if this is a reasonable approximation
and whether, instead, electron-positron pair production needs to be
properly taken into account in this scenario. In view of this, in what
follows we explore the significance of pair production during the
collapse of a rotating neutron star. In essence, we review the basic
mechanisms of pair production, as known from studies of pulsar
magnetospheres \citep{Harding2006}, concentrating in particular on the
photon-photon and photon-magnetic field mechanisms.

We recall that the occurence of photon-photon pair creation in a pulsar
depends sensitively on two fundamental parameters: the surface
temperature of the neutron star and the magnetic-field strength. Since
the blitzar model involves supramassive neutron stars that are thousands
to millions of years old, their surface temperature $T_s$ is 
expected to be well below $T_s < 10^6 \, K$ \citep{Chabrier2006}. For
such temperatures in the pair formation region, the field strength should
be $B > 0.1 \, B_{\rm cr}$ \cite{Harding2001}, where $B_{\rm cr}:= 4.4
\times 10^{13} \, {\rm G}$ is the so-called critical magnetic-field
strength. Thus, for an initial magnetic field of $\leq 4.4 \times 10^{12}
{\rm G}$, normally expected in a blitzar and the one considered in our
simulations here, photon-photon pair creation is strongly surpressed.

Another source of pair production is the photon--magnetic-field
mechaniscm, which involves the interaction of high energy photons with
strong magnetic field, which proceeds as follows. When in the exterior of
the pulsar a region is emptied of charges, thus creating a so-called
``gap'', an electric field parallel to the magnetic field develops as a result of
unipolar induction. This huge voltage drop across magnetic field lines is
capable of pulling charges from the neutron star surface and accelerate
them to high Lorentz factors. These accelerated charges, radiate
high-energy curvature photons, which, in turn, produce the
electron-positron pairs through a cascade interaction with the magnetic
field.

More specifically, the accelerated charges will attain a Lorentz factor
$\gamma$ that is proportional to the developed voltage drop $\Delta V$ and is given
by
\begin{align}
  \gamma= e\, \Delta V / m_e \, c^2 \, ,
  \label{eqn:gamma}
\end{align}
where $e$ and $m_e$ are the is the electric charge and the mass of the
electron, respectively. The emitted radiation will have a  characteristic
frequency given by 
\begin{align}
  \nu_c=\gamma^3 \frac{c}{r_c} \, ,
  \label{eqn:nu}
\end{align}
where $r_c$ is the radius of curvature of the magnetic-field line that
the charge will travel on. Electron-positron pairs are then created if
\citep{Sturrock1971, Ruderman1975} 
\begin{align}
  \gamma^3 \left(\frac{\hbar c}{2 m_e r_c c^2} \sin \theta\right)
  \left(\frac{B_{\rm loc}}{B_{\rm cr}} \right) 
  \simeq \frac{1}{15}   \, ,
  \label{eqn:pp}
\end{align}
where $B_{\rm loc}$ is the local magnetic field and $\sin(\theta)$ is the
``pitch'' angle between the photon and the magnetic-field line. It is
common to write $\sin\theta\simeq h/r_c$, where $h$ is the length of the
gap with the parallel electric field, (see, \eg \cite{Chen1993}).  Using
now Eqs. \eqref{eqn:pp} and \eqref{eqn:gamma}, we find the criterion for
triggering pair creation 
\begin{align}
 \Delta V < \Delta V_{\rm pp} \simeq \, 3 \times 10^{15}    &\left( 
 \frac{r_c}{20\, {\rm km}} \right)^{2/3}  
 \left( \frac{B_{\rm loc}}{10^{10}\, {\rm G}} \right)^{-1/3}  \nonumber \\
 & \left( \frac{h}{0.2\,{\rm km}}\right)^{-1/3} \quad {\rm statV }\, ,
 \label{eqn:volt}
\end{align}
or equivalently
\begin{align}
 E < E_{\rm pp} \simeq \, 1.5 \times 10^{11} & \left( 
 \frac{r_c}{20\, {\rm km}} \right)^{2/3}  
 \left( \frac{B_{\rm loc}}{10^{10}\, {\rm G}} \right)^{-1/3}  \nonumber \\
 & \left( \frac{h}{0.2\,{\rm km}}\right)^{-4/3} \quad {\rm statV/cm }\, .
 \label{eqn:electr}
\end{align}
In other words, no pair creation is expected from the interaction of
photons with the magnetic field as long as the voltage drop is below the
critical one $\Delta V_{\rm pp} \sim 3 \times 10^{15}\, {\rm
  statV}$. For a typical pulsar, the voltage drop can be estimated as
\begin{align}
\label{eq:DV_typ}
  \Delta V_{\rm typ} \sim 1.2 \times 10^{13} &\left( 
  \frac{B_{\rm loc}}{10^{12}\, {\rm G}} \right) \left( 
 \frac{\Omega}{1880\, {\rm rad/s}} \right) \nonumber \\
 & \left( \frac{h}{0.2\,{\rm km}}\right)^{2} \quad {\rm statV }\, ,
\end{align}
where $\Omega$ is the angular velocity of the star. 

While the estimate \eqref{eq:DV_typ} is simple to carry out for a
stationary pulsar, determining whether the voltage drop is always below
the critical one in a collapsing scenario, where all quantities in
Eq. \eqref{eqn:volt} change dynamically, is obviously more
complicated. In particular, during the collapse, the magnetic-field lines
are changing rapidly and the path of the accelerated charge is not
prescribed, but would need to be found self-consistently\footnote{Note
  that the light travel-time over a scale height of of $\sim 10\,{\rm
    km}$ is $\sim 30\, \mu{\rm s}$, which is longer than the typical
  timescale of variation of the magnetic-field lines.}. It follows that
the curvature radius will also vary dynamically and any reference scale
will not be valid but for a short time interval. Finally, within
microseconds the magnetic field in the region within which the
acceleration takes place may even change polarity, which means that the
path that the charge will follow might be chaotic, which questions the
efficiency of the curvature process.

Notwithstanding these caveats, in Sec. \ref{subsec:pair-cr} we will
follow the evolution of the parallel electric field in order to check
whether or not a sufficient voltage drop is created during the collapse
and hence whether pair production is at work. We can already anticipate
here that while pair production can take place during the collapse for
sufficiently large magnetic fields, this process is not efficient for the
typical values of the initial magnetic field in blitzars.

\begin{figure*}
  \begin{center}
    \includegraphics[width=0.9\textwidth]{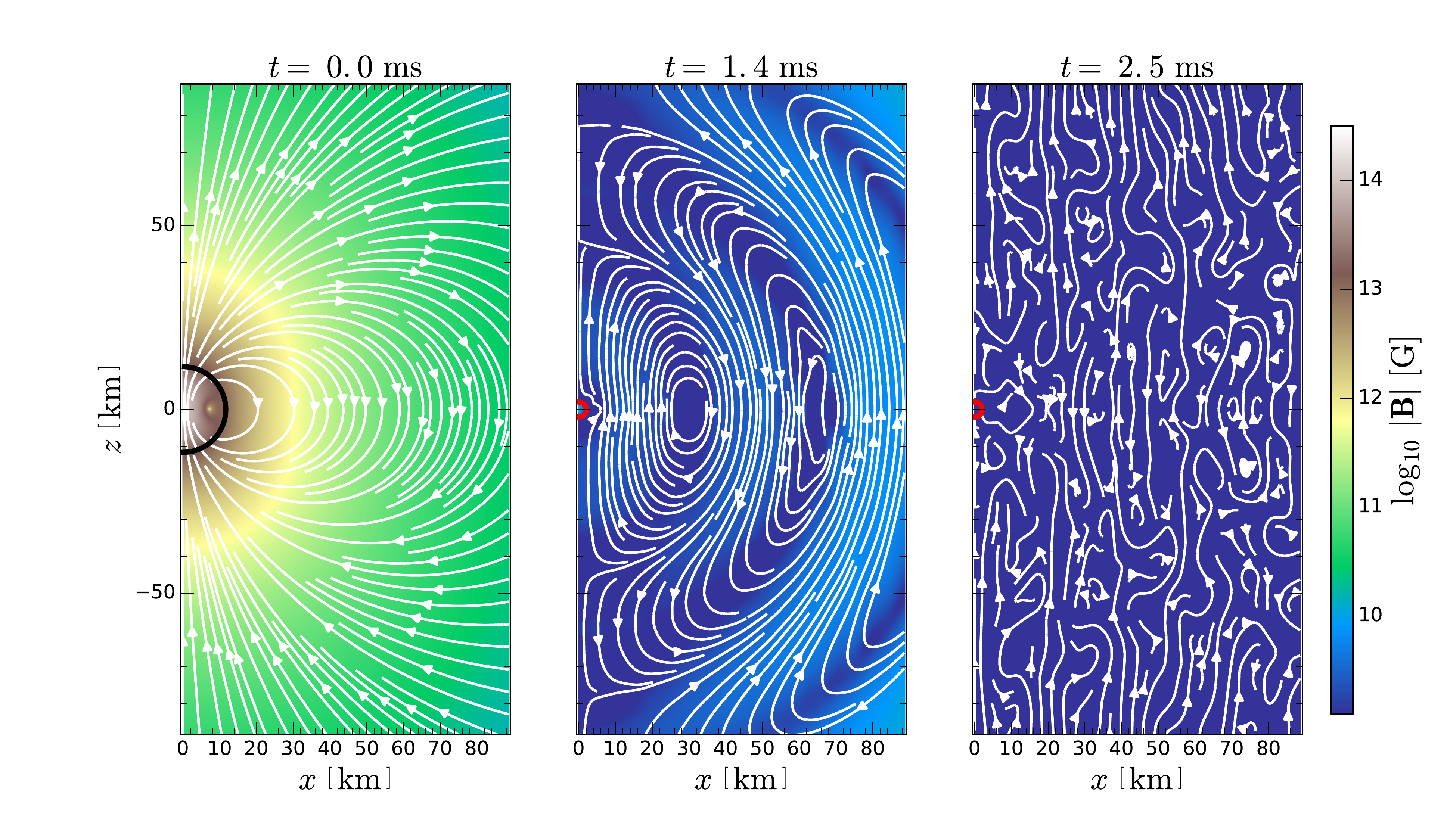}
  \end{center}
  \caption{Magnetic field strength $\left| \bm{B}\right|$ in the $(x,z)$
    plane shown with a colorbar and at three different times for the
    nonrotating model \ttt{F000.B13}. Also reported are the stellar
    surface (solid black line in the left panel), the apparent horizon
    (solid red line in the middle and right panels), and the
    magnetic-field lines (white lines). The initial magnetic field
    strength at the pole is $10^{13}\,{\rm G}$; note the lack of a final
    ordered magnetic field at late times (right panel) since the black
    hole produced is of Schwarzschild type.}
	\label{fig:B_norm_0}
\end{figure*}

\begin{figure*}
  \begin{center}
    \includegraphics[width=\textwidth]{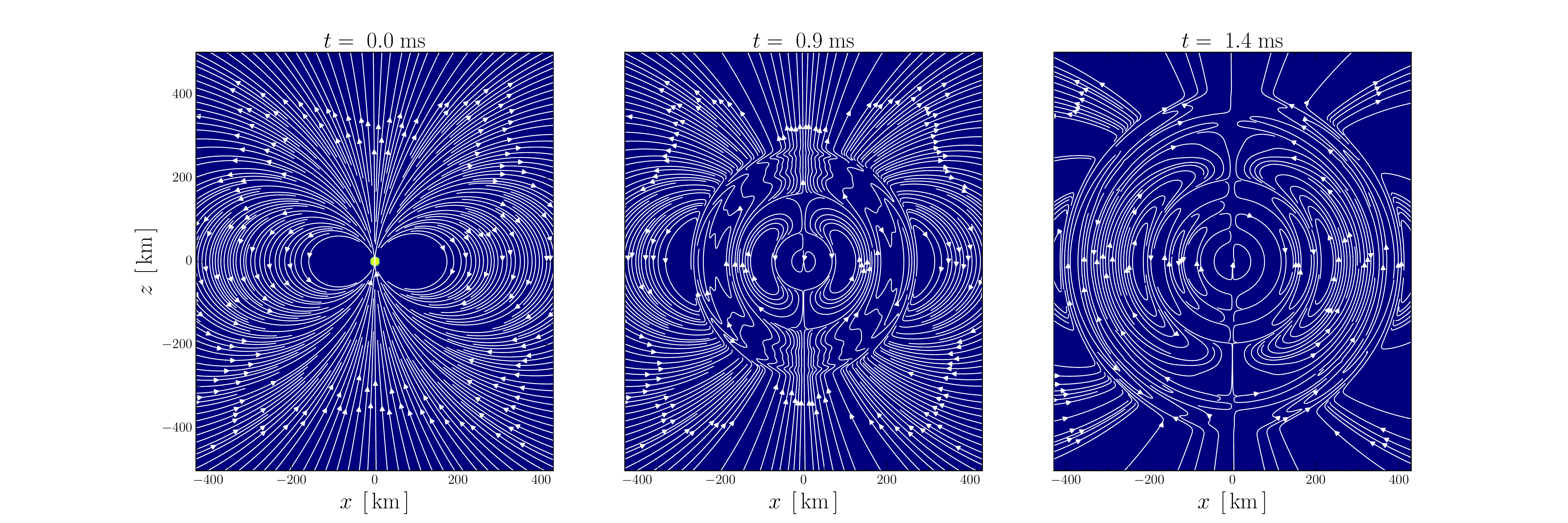}
  \end{center}
  \caption{Evolution of the magnetic-field lines in the $(x,z)$ plane for
    the same initial model \ttt{F000.B13} shown in
    Fig. \ref{fig:B_norm_0}, but presented here on a larger scale to
    highlight the global structure of the propagating EM wave.  The
    initial neutron star is indicated in green in the left panel as
    reference scale and the apparent horizon is not included in the
    middle and left panel, as it is too small for the scales considered.}
  \label{fig:B_norm_0_largescale}
\end{figure*}

\begin{figure*}
  \begin{center}
    \includegraphics[width=0.9\textwidth]{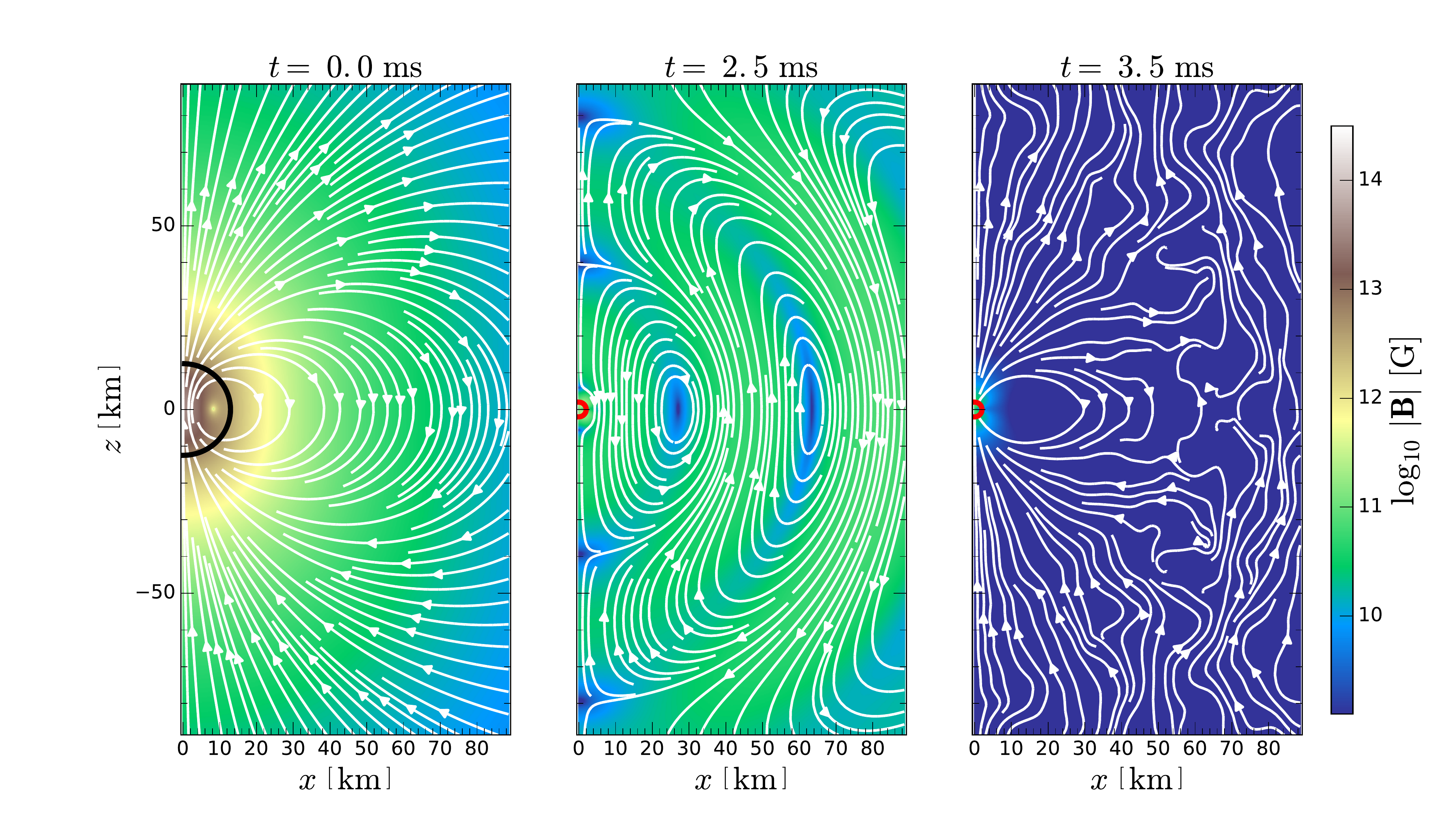}
    \includegraphics[width=0.9\textwidth]{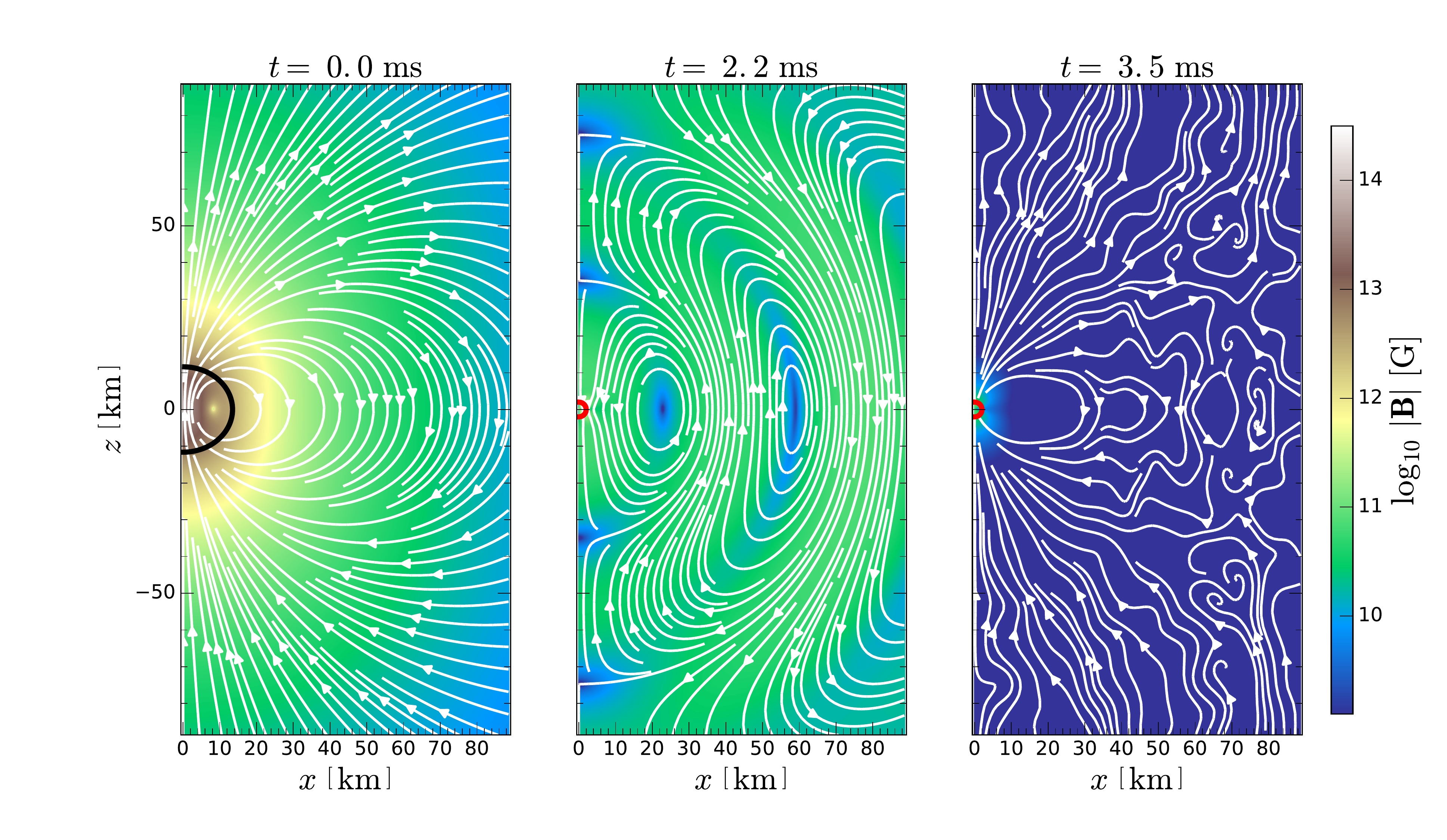}
  \end{center}
  \caption{\textit{Top row}: The same as in Fig. \ref{fig:B_norm_0}, but
    for the case of the initial model \ttt{F300.B13}, \ie a neutron
    star rotating at $300\,{\rm Hz}$ and with a magnetic field of
    $10^{13}\,{\rm G}$. Note the presence at late times (right panel) of
    an ordered magnetic since the black hole produced is of Kerr-Newman
    type. \textit{Bottom row}: The same as in Fig. \ref{fig:B_norm_0},
    but for the case of the initial model \ttt{F600.B13}.}
  \label{fig:B_norm_300}
\end{figure*}

\section{Numerical results}
\label{sec:nres}

\begin{figure}[t]
  \includegraphics[width=0.9\columnwidth]{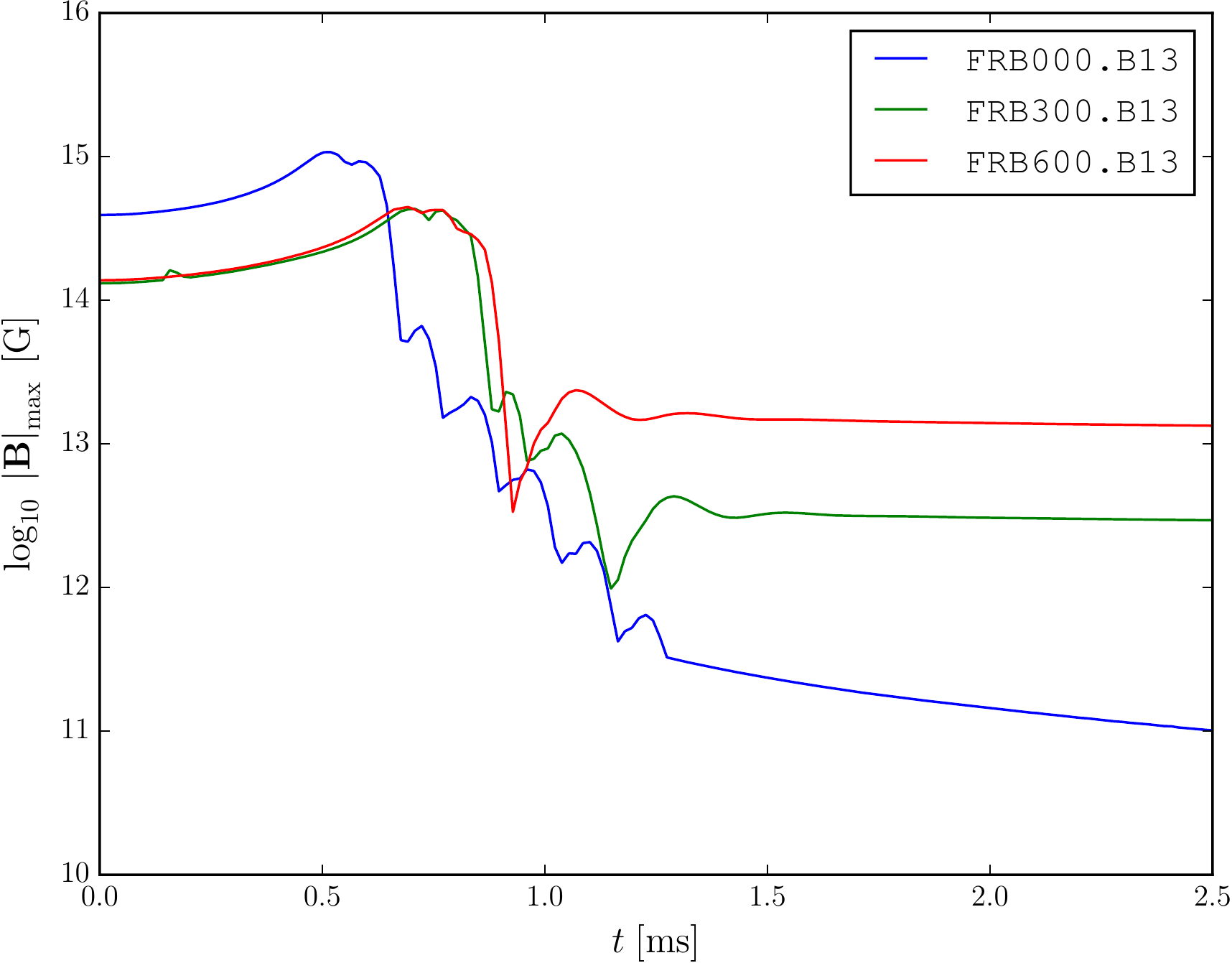}
  \hskip 1.0cm
  \includegraphics[width=0.9\columnwidth]{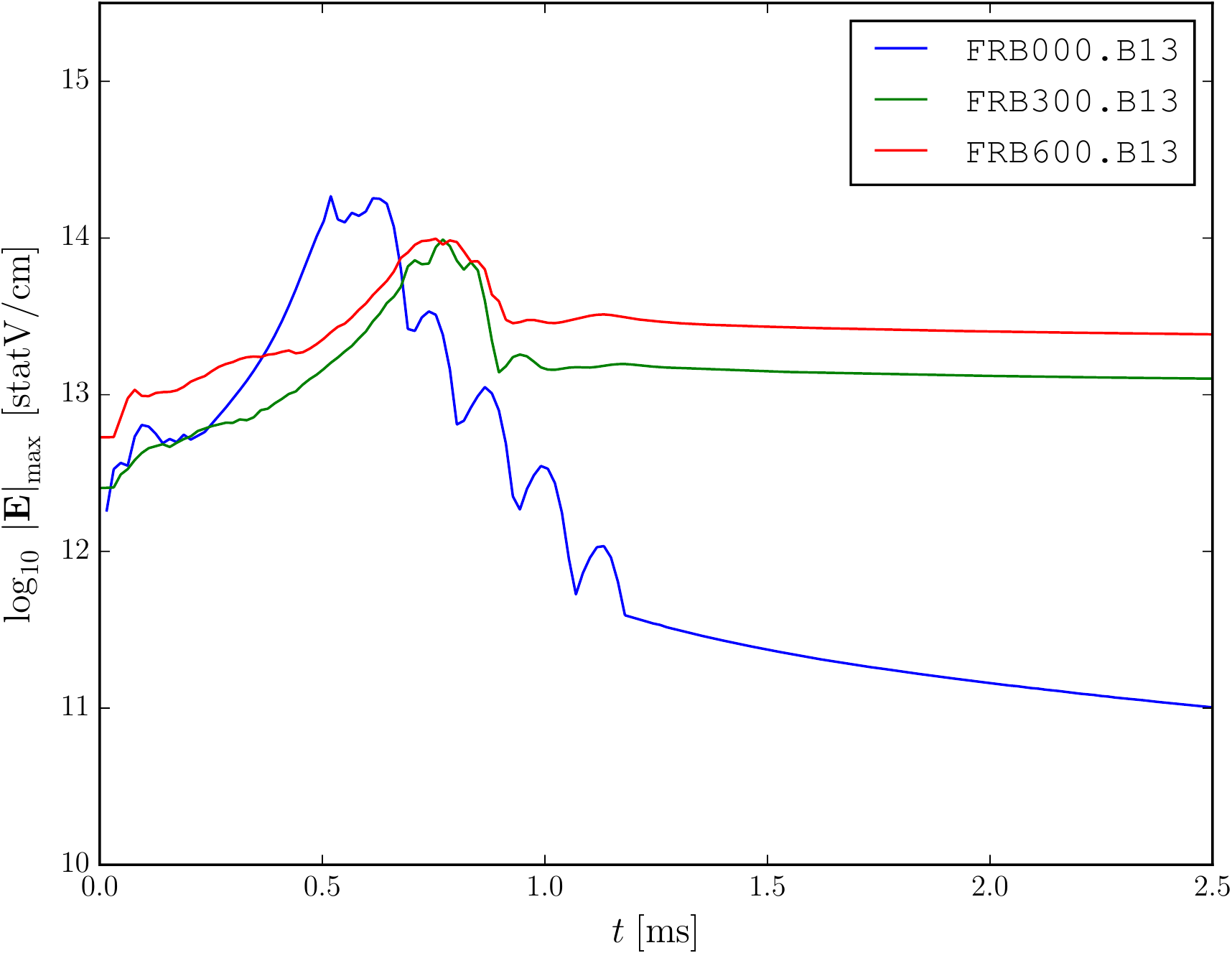}
  \caption{Maximum magnetic (left panel) and electric-field (right panel)
    strengths for selected nonrotating and rotating neutron stars models:
    \ttt{F000.B13}, \ttt{F300.B13}, and \ttt{F600.B13}.}
\label{fig:field_max}
\end{figure}

In what follows we will present the results of the simulations involving
the 17 neutron-star models that we have simulated.
We recall that the bulk properties of the matter dynamics have already
been studied in detail by several authors, starting from \citet{Font02c},
and more recently by \citet{Baiotti2007} and \citet{Liebling2010}. As shown
in detail in those works, the hydrodynamical collapse to a black hole
proceeds rapidly -- essentially on a dynamical timescale -- and does not
leave any remnant matter outside the apparent horizon, thus fully
justifying the assumption of an electrovacuum as the background over
which the electromagnetic waves emitted during collapse will propagate in
electrovacuum. Furthermore, since the overall phenomenological evolution
is very similar for all of the models described in Table
\ref{tab:initial}, we will discuss in more detail only the results for
the models \ttt{F000.B13}, \ttt{F300.B13} and \ttt{F600.B13} as
they are three representative of three qualitatively different
behaviours. More specifically, we will next first discuss the dynamics of
the magnetic field during the collapse (Sec. \ref{sec:B_dynamics}) and
subsequently the properties of the EM emission and the magnitudes of the
energy losses (Sec. \ref{sec:em_emission}).

\subsection{Magnetic-field dynamics}
\label{sec:B_dynamics}

Although the nonrotating model \ttt{F000.B13} was already considered
by \citet{Dionysopoulou:2012pp}, we will briefly discuss it here as it
provides a useful reference solution. We recall that the neutron star is
initially endowed with a dipolar magnetic field, as can be seen in the
first panel of Fig. \ref{fig:B_norm_0}.  When the collapse begins, a
strong discontinuity is produced in the magnetosphere as the whole
surface of the star suddenly starts to move inwards. This ``magnetic
shock'' propagates outwards at the speed of light, reaching the at $\sim
300\,{\rm km}$ in almost $\sim 1\,{\rm ms}$ and essentially destroys the
dipolar field structure (see middle panel of Figs. \ref{fig:B_norm_0},
\ref{fig:B_norm_0_largescale}). Behind this shock, and when the apparent
horizon is formed, the magnetic-field lines are violently snapped. At
this point, quadrupolar EM radiation is produced and the EM fields
propagate outwards essentially as EM waves in electrovacuum. At the same
time, the electric and magnetic fields near the stellar surface and the
apparent horizon constantly decay, losing any ordered large-scale
structure as they cannot be sourced by the emerging Schwarzschild black
hole (see right panel of Fig. \ref{fig:B_norm_0}). Figure
\ref{fig:B_norm_0_largescale} provides essentially the same information
reported by Fig. \ref{fig:B_norm_0}, but it shows it on a larger scale of
$\sim 450\,{\rm km}$ so as to highlight the coherent large-scale
structure of the magnetic field and the quadrupolar nature of the emitted
EM radiation.

What cannot be shown in detail in Figs. \ref{fig:B_norm_0_largescale} and
\ref{fig:B_norm_0} are the magnetic-field properties near the apparent
horizon. A careful analysis of the dynamics of the magnetic-field lines
reveals that when the apparent horizon is formed, magnetic-field lines
still pass through it. However, as the black hole starts to ringdown,
magnetic loops are generated right outside the horizon and then propagate
outwards. At this point in time, no magnetic-field lines passes through
the horizon and the magnetic-field lines strength has decreased
considerably.

Although what described above refers to a nonrotating model, the overall
magnetic-field evolution is quite similar, at least on large scales, also
for the rotating ones. This is shown in Fig. \ref{fig:B_norm_300}, which
is the same as in Fig. \ref{fig:B_norm_0}, but refers in the top row to
the initial model \ttt{F300.B13}, \ie a neutron star rotating at
$300\,{\rm Hz}$ and with a pole magnetic field of $10^{13}\,{\rm G}$,
while in the bottom row it refers to the initial model \ttt{F600.B13}\footnote{We 
are considering very high spin rates in order to explore the variation 
of the released EM energy as a function of the stellar rotation. However, 
typical values for the spin frequency in the blitzar scenario are  
of a few Hz only.}.
Also in these cases, in fact, as the collapse begins and the
neutron-star's surface starts to shrink, the magnetosphere is disrupted
and a magnetic shock is produced by the snapped magnetic field
lines. Again, a quadrupolar EM radiation is produced near the black hole,
which propagates outwards as a travelling EM wave. Note that in the case
of rotating collapsing stars, all magnetic loops that are formed near the
apparent horizon actually pass through it and are sourced by some
current below the apparent horizon.

Another important difference between the rotating and nonrotating models
is the different late-time magnetic-field dynamics close to the black
hole. This can be appreciated by comparing the right panel of
Fig. \ref{fig:B_norm_0} with the corresponding right panels of
Fig. \ref{fig:B_norm_300}. While in fact in the first case the magnetic
field is unstructured at all scales, in the second cases the magnetic
field exhibits a clear dipolar structure whose strength depends on
the initial rotation of the collapsing star: a higher rotation rate
yields a higher asymptotic magnetic field at the horizon. This is shown
in the left panel of Fig. \ref{fig:field_max}, which reports the maximum
magnetic-field strengths for the selected nonrotating and rotating
neutron stars models: \ttt{F000.B13}, \ttt{F300.B13}, and
\ttt{F600.B13}.

The reason behind this different behaviour is simple and has to be found
in the fact that the rotating models have an initial charge
induced by the nonzero electric field. Indeed, although our prescription
for the electric field, \ie a corotating interior electric field matched
to a divergence-free electric field produced by a rotating magnetised
sphere, is the one that minimises the induced charge, our rotating
neutron star models are electrically charged initially. As a result, their
gravitational collapse will not lead to Kerr black holes but, rather, to
Kerr-Newman black holes \citep{Nathanail2017}. 

Since we do not model any additional process that would change the net
charge of the system, \eg via pair creation, the initial charge of the
neutron star is essentially all conserved and is acquired by the black
hole. Such a charge $Q$ is only a very small fraction of the mass of the
black hole $M_{_{\rm BH}}$, \ie $Q \sim 10^{-4}\,M_{_{\rm BH}}$, even for
the highest rotation model \citep{Nathanail2017}, but it leaves the black
hole with EM fields that could be astrophysically significant. 

The evolution of the electric field for the three representative cases is
shown in the right panel of Fig. \ref{fig:field_max} and has a behaviour
that is very similar to that of magnetic field (left panel).  However, it
should be borne in mind that the net charge measured is effectively very
small and at the limit of the numerical accuracy of our simulations (we
recall that our highest spatial resolution is $h=0.1\,M_{\odot}$ at
most). In reality, however, if such a collapse would take place in an
astrophysical scenario, then the abundant free charges that accompany
astrophysical plasmas would neutralise it very rapidly, yielding
therefore a standard Kerr solution.

\citet{Dionysopoulou:2012pp}, but also \citet{Baumgarte02b2} and
\citet{Lehner2011}, computed the late time evolution of these EM fields
in terms of the magnetic flux across a given surface and showed that they
decay exponentially, following the ringdown of the newly formed
nonrotating black hole. This behaviour has been reproduced by our
simulations and can be seen in Fig. \ref{fig:field_max}, both for the
magnetic (left panel) and for the electric field (right panel). Hence,
and as remarked by \citet{Falcke2013}, should the EM emission from an FRB
be accompanied by an exponentially decaying EM signal, it would provide
unambiguous evidence that a black hole has indeed been produced together
with the FRB (see also Fig. \ref{fig:Poynting_comp} and related
discussion).

\subsection{Electromagnetic-energy emission}
\label{sec:em_emission}

Having established that the collapse of a magnetised neutron star, be it
nonrotating or rotating, leads to a magnetospheric destruction and to
the production of an intense emission of EM waves, we will next discuss
the energetics and the typical duration of these signals so as to compare
the results of our simulations with the phenomenology associated with
FRBs.

In particular, we compute the EM luminosity generated during the collapse
through the expression
\begin{align}
  L_\mathrm{EM} := 
  \oint_{\Sigma} \boldsymbol{S}_{_{\mathrm{EM}}}\cdot \, d \mathbf{\Sigma} \,,
  \label{eqn:poynting}
\end{align}
on a spherical coordinate surface $\Sigma$ at a radial distance $r\simeq
205\,{\rm km}$ from the collapsing neutron star, where
$\boldsymbol{S}_\mathrm{EM}$ is the Poynting vector.

\begin{figure*}
  \begin{center}
    \includegraphics[width=0.925\columnwidth]{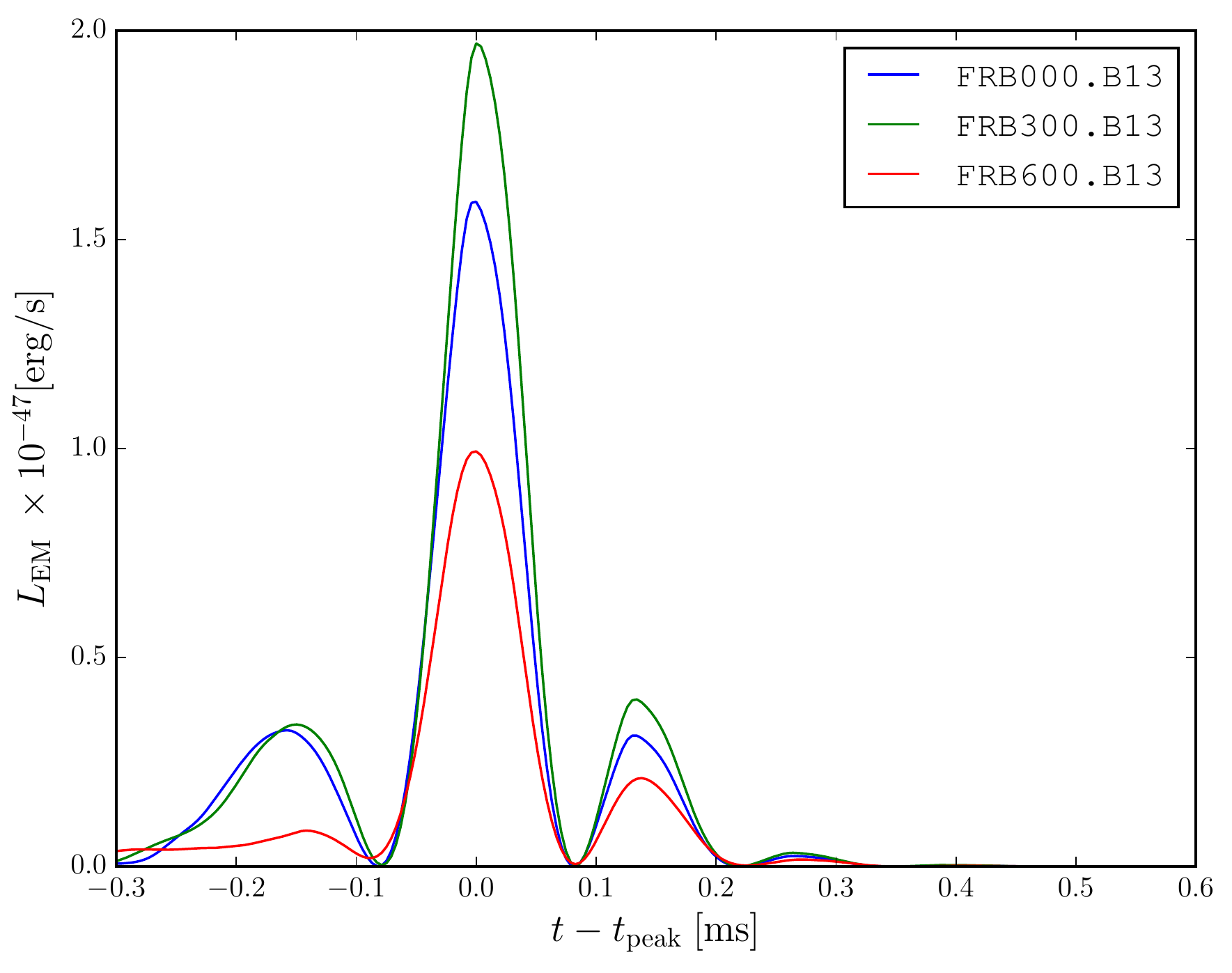}
    \hskip 1.0cm
    \includegraphics[width=0.9\columnwidth]{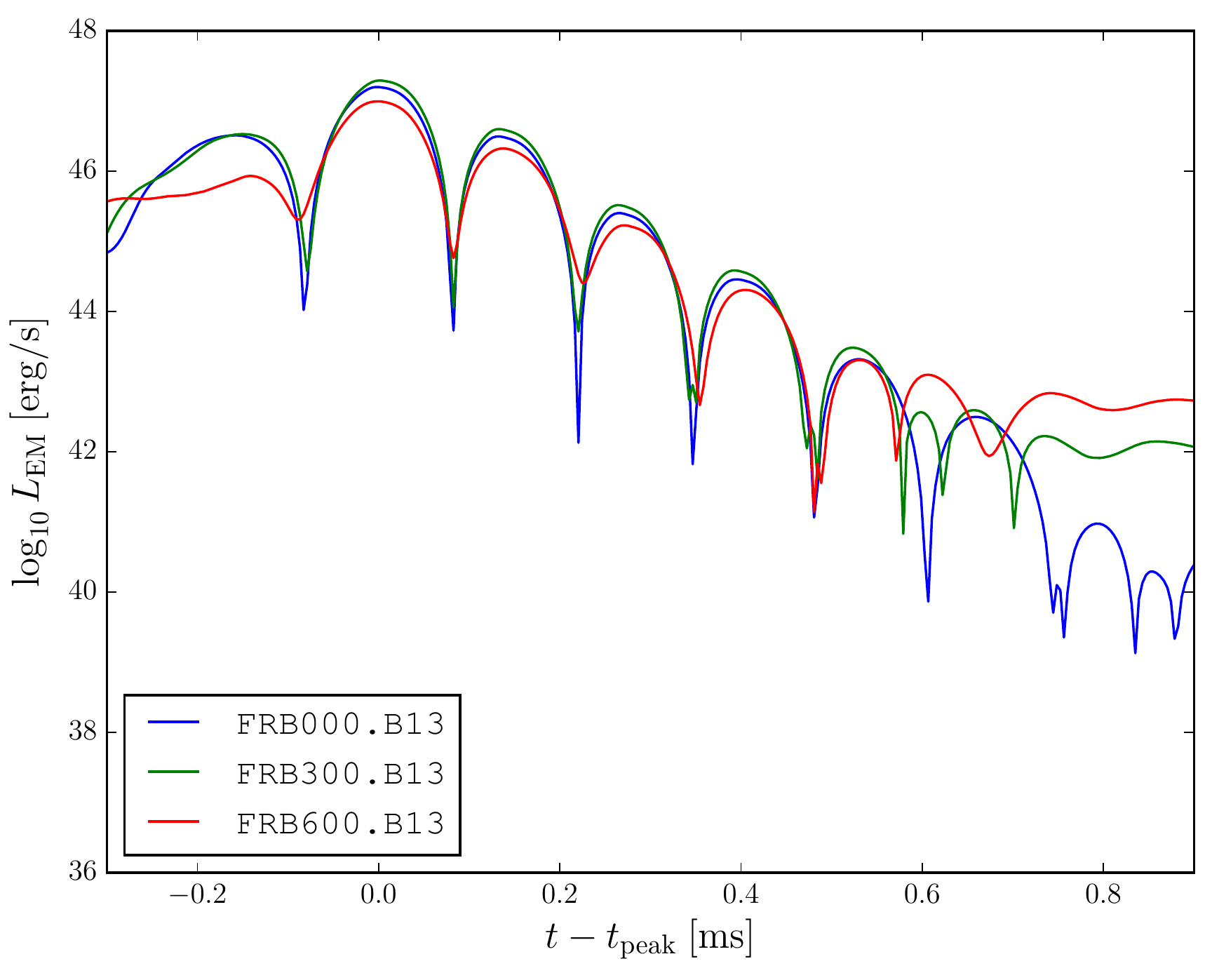}
  \end{center}
  \caption{EM luminosity for three representative models
    \ttt{F000.B13}, \ttt{F300.B13}, and \ttt{F600.B13} as
    extracted at $205\,{\rm km}$. The different signals in retarded time
    are aligned so that they coincide when the largest peak reaches the
    detector. The two panels report the same data but show it either on a
    linear scale (left panel) or in a logarithmic one. Note that the
    small differences in the initial magnetic field strengths at the
    poles, see Tab. \ref{tab:initial}, have been scaled out assuming a
    $B^2$ scaling.}
  \label{fig:Poynting_comp}
\end{figure*}
\begin{figure*}
  \begin{center}
    \includegraphics[width=0.95\textwidth]{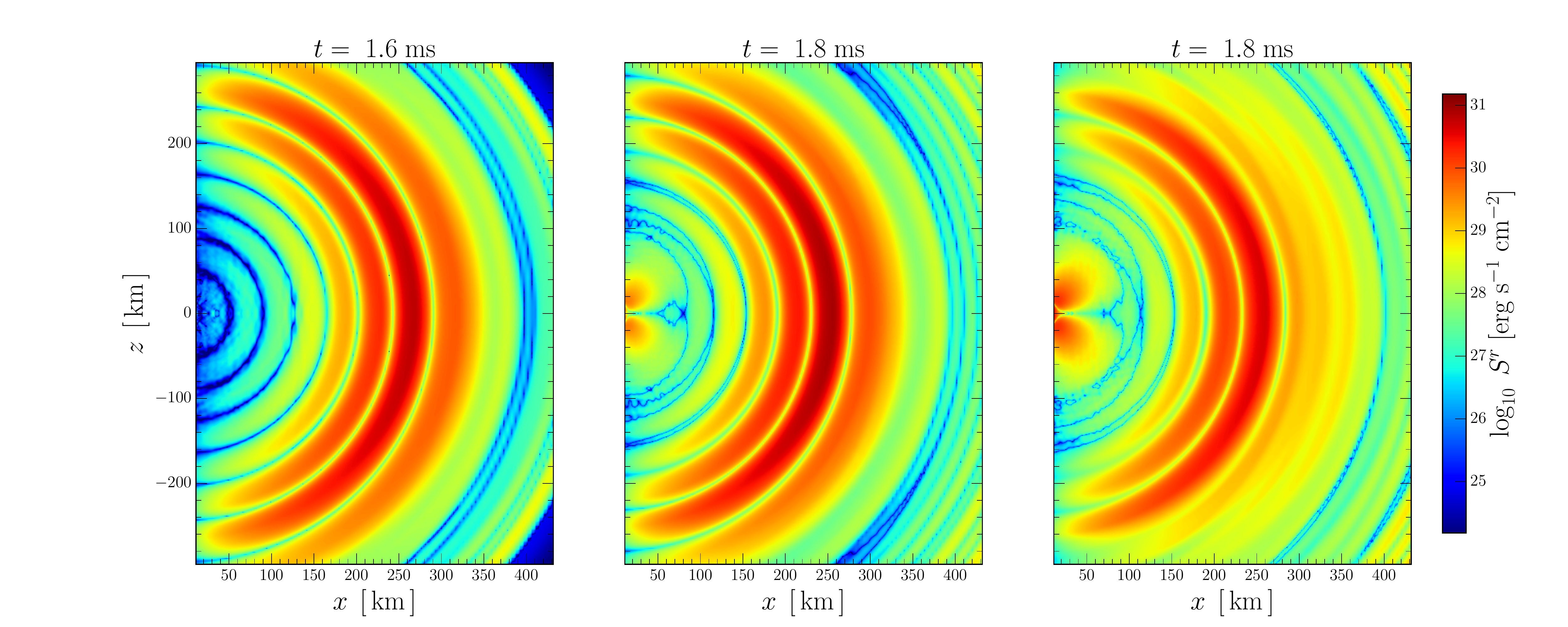}
  \end{center}
  \caption{Radial component of the Poynting vector $\boldsymbol{S}$ in
    the $(x,z)$ plane for models \ttt{F000.B13}, \ttt{F300.B13}, and
    \ttt{F600.B13}. It is this complex structure that leads to the
    multi-peaked EM emission reported in
    Fig. \ref{fig:Poynting_comp}. Note that the small time difference
    between the models is a result of the slightly delayed collapse for
    fast rotating models. Note that the small differences in the initial
    magnetic field strengths at the poles, see Tab. \ref{tab:initial},
    have been scaled out assuming a $B^2$ scaling.}
\label{fig:2D_Poynting}
\end{figure*}

Figure \ref{fig:Poynting_comp} reports the computed luminosity
\eqref{eqn:poynting} as a function of time for three representative
models in a linear (left panel) and in a logarithmic scale (right panel),
respectively. The signals from the different stellar models are expressed
in retarded time and are aligned so that they coincide when the largest
peak reaches the detector.  Clearly, all of the luminosity curves show a
well defined and dominating sub-millisecond pulse, in close analogy with
the observations of FRBs \citep{RaneLorimer2017}. Furthermore,
the main pulse is always accompanied by both a precursor that is about
$10\%$ smaller and then by a successive pulse that is of similar
amplitude (\cf left panel of Fig. \ref{fig:Poynting_comp}).
Interestingly, this pattern of peaks is rather similar to the one
observed for FRB 121002 \citep{Champion2016}, thus highlighting that a 
blitzar model can accommodate rather
naturally the multi peaked phenomenology of FRBs. Furthermore, and as
discussed earlier, even when the black hole is formed, the EM emission
does not cease and the black hole rings down shedding its EM
perturbations in terms of a wave-train of EM pulses (\cf right panel of
Fig. \ref{fig:Poynting_comp}). It is exactly the detection of this
ringing-down signature that would corroborate the blitzar model as the
most plausible one to describe non-repeating FRBs.

Figure \ref{fig:Poynting_comp} also allows us to deduce two important
results that will be further discussed also in the following. Firstly,
the overall EM energy radiated in \tbf{the} whole collapse depends only
very weakly on the stellar rotation rate (indeed, the radiated energy
differs only of $ 30\%$ when going from the nonrotating model to the most
rapidly rotating model considered, having scaled out the small
differences in the initial magnetic field assuming a $B^2$ scaling.).
Secondly, the timescale for the EM emission is comparable in all cases
and $99.99\%$ of the energy is emitted within one millisecond; this
result will be used later on when estimating an expression for the
radiated energy.

It is possible to appreciate the multi-peaked structure of the EM
emission from the collapsing star through the two-dimensional section on
the $(x,z)$ plane of the radial component of the Poynting vector
$\boldsymbol{S}$ and appears in the integral \eqref{eqn:poynting} for the
EM luminosity. This is shown in Fig. \ref{fig:2D_Poynting}, where all the
pulses are visible and distinct as they travel outwards. Considering that
the colorcode reports the Poynting vector in a logarithmic scale and that
the pulses move at the speed of light, it is to reconstruct from
Fig. \ref{fig:2D_Poynting} both the precursor and the exponentially
decaying structure of the EM luminosity shown in
Fig. \ref{fig:Poynting_comp}. Figure \ref{fig:2D_Poynting} also
highlights the quadrupolar nature of the EM emission, with most of the
intensity concentrated near the equatorial plane of the rotating
star. This lack of anisotropy has direct consequences on the event rate
of blitzars and detection rate of FRBs, indicating that if blitzars are
responsible for FRBs, then the event rate should be close to a factor of two
larger than the detection rate. As a final remark, we should point out 
that the correct event rate of blitzars would be determined by simulating 
realistic pulsars, meaning that the rotational axis is misaligned with the 
magnetic dipole moment. We intent to extent our present work to the 
misaligned case.

Having described the overall energetic of the EM emission, it is
interesting to correlate the measured radiated energy with the basic
properties of the stellar models, namely, the magnetic-field strength and
the rotation rate. The ultimate goal is to derive a phenomenological
expression that would provide a simple estimate of such quantities on the
basis of the measured energetics of the observed FRB. \tbf{Hence}, we
compute the radiated energy simply as the time integral of the EM
luminosity, \ie
\begin{equation}
E_{_{\rm EM}} := \int L_{_{\rm EM}}(t)\,dt \,,
\end{equation}
and report in Table \ref{tab:initial} the values computed for all of the
different models. A rapid look at the table shows that this radiated
energy is effectively almost constant across all models and that the
radiated EM energy is only weakly dependent on the rate of rotation of
the star. This behaviour is rather different from the corresponding
energy radiated in GWs $E_{_{\rm GW}}$. While the two energies are indeed
comparable, \ie $E_{_{\rm GW}} \approx E_{_{\rm EM}} \approx
10^{43}\,{\rm erg}$, for a collapsing neutron star with initial magnetic
field $B_{\rm pol} \approx 10^{13}\ \mathrm{G}$ and rotation frequency $f_{\rm
  spin} \approx 100\,{\rm Hz}$, the radiated GW energy has been shown to
depend steeply on the dimensionless angular momentum of the star
$\tilde{J} := J/M^2$ and, in particular, to follow a relation of the type
$E_{_{\rm GW}} \propto \tilde{J}^4$ for rotation rates almost up to the
mass-shedding limit \citep{Baiotti2007}. This difference, however, is not
surprising and is to be found in the fact that while the EM energy
radiated reflects the actual energy stored in the magnetosphere, which
does not vary significantly with rotation, the GW energy depends on a
high time derivative of the quadrupole moment and is therefore much more
sensitive to the variations of the latter with the spin rate.

Next, we take the phenomenological expression proposed by
\citet{Falcke2013}, for the available power in the magnetosphere of a
typical pulsar [\cf Eq. (4) of \citet{Falcke2013}]
\begin{equation}
\label{eq:powerMS}
P_{_{\rm MS}} \simeq 8.4 \times 10^{44} \;
\eta_{_{\rm B}} \, t_{\rm ms}^{-1}\, b_{12}^2 \, r_{10}^{3}
\ \ {\rm erg\ s}^{-1}\,,
\end{equation}
where $\eta_{_{\rm B}}$ is the magnetic-energy efficiency, that is, the
fraction of magnetic energy in the magnetosphere that is effectively
dissipated, $\Delta t = t_{\rm ms}\,1 {\rm ms}$ is the duration of the 
burst, while $b_{12}$ and $r_{10}$ are the magnetic field of the
star and its radius in units of $10^{12}\,{\rm G}$ and $10\,{\rm km}$,
respectively, \ie $B_{\rm pol} =: b_{12}\,10^{12}\,{\rm G}$ and $R=:
r_{10}\,10\,{\rm km}$. Note that although $\eta_{_{\rm B}}$ is unknown
(but see below), a value of order unity already provides a value for the
luminosity that is in very good agreement with the one observed in FRBs. 
Here after, we will refer to the magnetic-energy efficiency as
$\eta_{_{\rm Bev}}$, since all our results were obtained assuming 
an electro-vacuum. 

\begin{figure}
  \begin{center}
    \includegraphics[width=1.0\columnwidth]{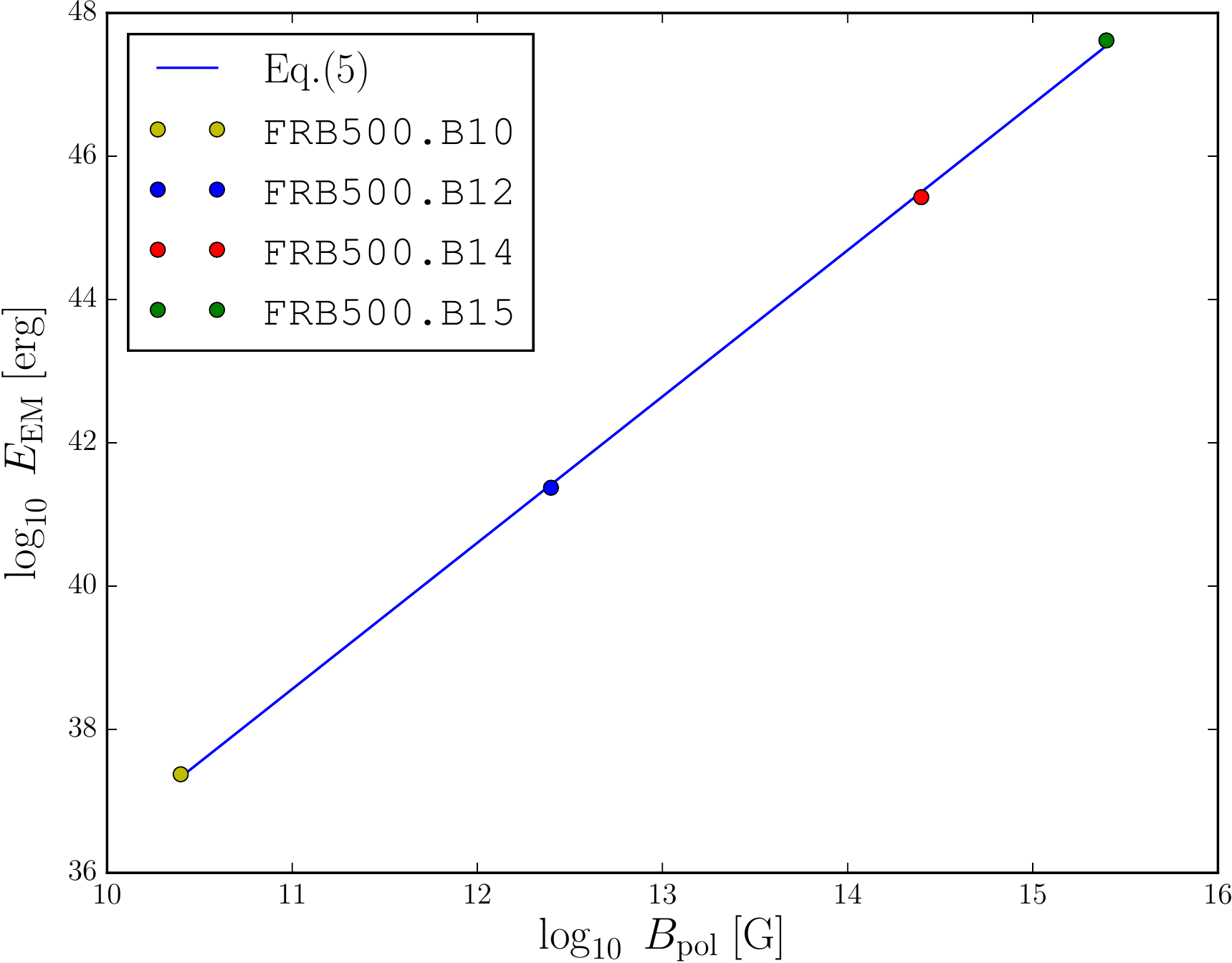}
  \end{center}
  \caption{Scaling of the emitted EM energy extracted at $205\,{\rm km}$
    with the initial value of the magnetic field on the pole $B_{\rm
      pol}$.  The models shown are \ttt{FRB500.B10}
    -\ttt{FRB500.B15}.}
    \label{fig:scaling}
\end{figure}

Note that expression \eqref{eq:powerMS} assumes a quadratic scaling on
the initial magnetic field; while this is reasonable from an energetic
point of view, it remains an assumption. However, it can be easily
verified by computing the energy emission when considering initial
stellar models with the same spin frequency but different degree of
magnetisation, \ie, in terms of the initial models
\ttt{F500.B10}--\ttt{F500.B15}. The results of this calculation are
shown in Fig. \ref{fig:scaling}, which reports the emitted energy
$E_{_{\rm EM}}$ extracted at $205\,{\rm km}$ 
as a function of the initial value of the magnetic field at the 
pole $B_{\rm pol}$. The log-log plot clearly shows that there is a
power scaling between $E_{_{\rm EM}}$ and $B_{\rm pol}$ and a fitting procedure
shows that the scaling exponent is indeed $2.04\pm 0.02$, as predicted by
\citet{Falcke2013}. 

The data in Fig. \ref{fig:scaling} also allows us to fix the
magnetic-energy efficiency $\eta_{_{\rm Bev}}$. As mentioned above, in
fact, the timescale for the EM emission is essentially independent of the
initial stellar rotation rates, at least for the rates considered here
(\cf Fig. \ref{fig:Poynting_comp}), and is of the order of one
millisecond, \ie $\Delta t_{_{\rm EM}}/{\rm ms} = 1 = t_{\rm ms}$. As a
result, we can express the emitted energy as
\begin{equation}
\label{eq:scaled_EEM}
E_{_{\rm EM}} = P_{_{\rm MS}}\, \Delta t_{_{\rm EM}}
\simeq 8.4 \times 10^{41} \;
\eta_{_{\rm Bev}} \, b_{12}^2 \, r_{10}^{3}
\ \ {\rm erg}\,.
\end{equation}                          
Using expression \eqref{eq:scaled_EEM} and the data in
Fig. \ref{fig:scaling} we therefore deduce via the quadratic fit that the
magnetic-energy efficiency is $\eta_{_{\rm B}} = 1.8 \%$ as for the
computed models \ttt{F500.B10}--\ttt{F500.B15}. We note that
although the efficiency $\eta_{_{\rm Bev}}$ is only weakly dependent on the
initial stellar rotation rate, it is not totally independent of
it. Repeating similar calculations also for all stellar models in Table
\ref{tab:initial} reveals that the higher efficiency of $\eta_{_{\rm Bev}}
= 3.6 \% $ is for the model with spinning frequency of $f_{\rm spin} =
100\, {\rm Hz}$ and the lowest one $\eta_{_{\rm Bev}} = 1.4\% $ for the
model with spinning frequency of $f_{\rm spin} = 400\, {\rm Hz}$. This
variance is not unexpected as the dependence of $\eta_{_{\rm Bev}}$ on
$f_{\rm spin}$ is weak and it is well known that the dynamics of the
collapse ``slows down'' as the spin rate of the neutron stars increases
\citep{Baiotti2007}. Given this variance, we compute an average value of
$\eta_{_{\rm Bev}} = (2.1 \pm 0.5) \%$ and hence obtain a phenomenological
expression for the EM power released by a blitzar as given by
\begin{equation}
\label{eq:powerMS_2}
P_{_{\rm MS}} \simeq 1.7 \times 10^{43} \;
t_{\rm ms}^{-1}\, b_{12}^2 \, r_{10}^{3}
\ \ {\rm erg\ s}^{-1}\,,
\end{equation}
while the corresponding energy is
\begin{equation}
\label{eq:scaled_EEM_2}
E_{_{\rm EM}} \simeq 1.7 \times 10^{40} \; b_{12}^2 \, r_{10}^{3}
\ \ {\rm erg}\,.
\end{equation}
Within the blitzar model, therefore, once an FRB of a given energy is
measured, using Eqs. \eqref{eq:powerMS_2} and \eqref{eq:scaled_EEM_2} it
is possible, at least in principle, to set constraints on either the
radius of the collapsing star or on its magnetic field.

\begin{figure*}
\begin{center}
  \includegraphics[width=0.9\textwidth]{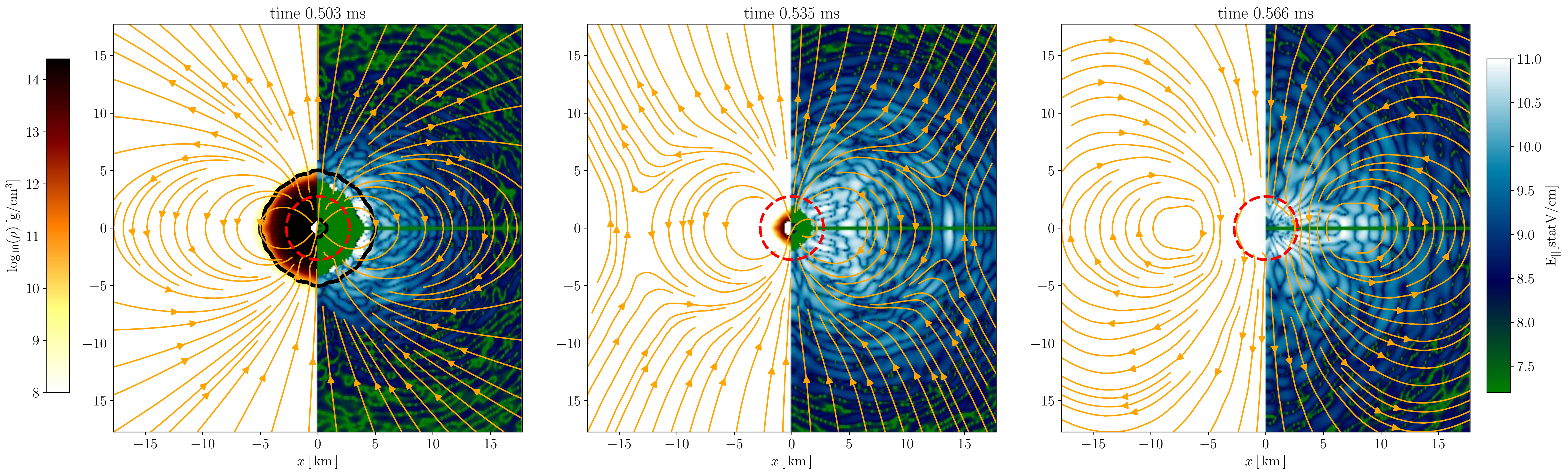}
\end{center}
\caption{Rest-mass density (left pnels) and electric field parallel to
  the magnetic field (right panels) in the $(x,z)$ plane shown at three
  different times for model \ttt{F001.B12}. Also reported are the stellar
  surface (solid black line in the left panel), the apparent horizon
  (dashed red line), and the magnetic-field lines (orange lines).}
\label{fig:EB}
\end{figure*}

We should remark that the considerations made so far are simply
``bolometric'' in the sense that we are simply computing the EM energy
emitted from the collapsing process in terms of the Poynting flux
measured at large distances from the source. In this respect, we have not
at all discussed how this bulk energy is then channelled, most likely in
a coherent manner, to produce the observed radio emission in FRBs. A
simplified curvature-radiation model that uses blitzar emission to
reproduce the observed FRB phenomenology is discussed by
\citet{Falcke2013} and we still consider it a reasonable first radiation
model for blitzars. Our future intention is to couple our present
simulations with a curvature-radiation model, in order to produce
realistic radiation imprints of a blitzar. At the same time, we refer the
interested reader to the recent works of \cite{Katz2014, Kumar2017} and
\citet{Ghisellini2017}.

\subsection{On the pair production in a blitzar scenario}
\label{subsec:pair-cr}

As anticipated in Sec. \ref{sec:pair-pro}, it is important to assess the
occurrence of pair production under the physical conditions that are
produced during a blitzar scenario. \tbf{To this purpose}, we follow
numerically the evolution of the maximum value of the parallel electric
field, $\boldsymbol{E}_{||} := \boldsymbol{E} \cdot
\boldsymbol{B}/|\boldsymbol{B}|$ responsible for any particle
acceleration and hence pair production. Our reference model is that of a
typical supramassive neutron star involved in a blitzar scenario, namely,
a star with the magnetic field is $B_{\rm pol} = 10^{12}\,{\rm G}$ and
period of $1\, {\rm sec}$ ($f_{\rm spin} = 1\,{\rm Hz}$) With such
initial magnetic field and rotation rate, the star is supposed to have
passed its death line, where no pair creation and pulsar emission is
expected to take place \citep{Chen1993}.

In Fig. \ref{fig:EB} we show the evolution of the rest-mass density (left
portions of the panels) and of the maximum of the parallel electric field
(right portions of the panels) at three representative times a typical
evolution: one just before all matter is lost inside the black hole and
two shortly afterwards; these are also the times when
$\boldsymbol{E}_{||}$ reaches its highest value. Also shown in
Fig. \ref{fig:EB} are the stellar surface (solid black line in the left
panel), the apparent horizon (dashed red line), and the magnetic-field
lines (orange lines).

Note that as the collapse proceeds, the parallel electric field grows,
but also that the largest values are confined within the star and below
its surface. Indeed, the parallel electric field remains below the
critical value for pair creation $E_{\rm pp}$ [\cf Eq. \eqref{eqn:electr}]. The
maximum of this growth takes place shortly before all matter is lost
behind an apparent horizon, so that there is only a very short window in
time, \ie of the order of a fraction of a microsecond, during which
charges could be pulled from the stellar surface.

In summary, the results presented in Fig. \ref{fig:EB} show that the
typical strength of the magnetic field in a blitzar scenario, \ie $B_{\rm
  pol} = 10^{12}\,{\rm G}$, is at the limit of the physical conditions
below which pair creation is strongly suppressed. This finding provides
us with the confidence on the robustness of the results presented here in
a pure electrovacuum scenario. On the other hand, the results in
Fig. \ref{fig:EB} also indicate that for \tbf{larger} initial
magnetic-fields, pair creation is very likely to take
place. \tbf{Interestingly, if pair creation does take place during the
  collapse, the electromagnetic emission is likely be different, hence
  providing an important signature for the occurrence of the pair
  creation. More specifically, it is reasonable to assume that together
  with the emission discussed so far and due to the global snapping of
  the magnetic-field lines, photons produced from the pair cascade, and
  that do not have sufficient energy to further pair create, could then
  diffuse through the stellar exterior leading to an additional
  emission. This scenario, which could be considered a \emph{``dirty
    blitzar''}, would then have a multi-frequency radiation spectrum. A
  more detailed study is necessary to further explore this speculation,
  both in the theoretical modelling and in the analysis of the
  observational data.}

\section{Conclusions}
\label{sec:conclusions}

Understanding the physics of astronomical systems dominated by extreme
gravity and ultra-strong magnetic fields is at the heart of high-energy
astrophysics. In this context, the collapse of a rotating and magnetised
neutron star represents a perfect example,. which has been explored via
numerical simulations in full general relativity by several authors in
the recent past \citep{Baumgarte02b2, Lehner2011,
  Dionysopoulou:2012pp}. In addition to the physical insight that these
investigations have brought, they have been also useful to define a
theoretical framework that provide a simple explanation of some of the
most exciting and yet mysterious astronomical objects that have been
recently observed: fast radio bursts. The blitzar model, in fact,
involves the gravitational collapse of a rotating and magnetised neutron
star and has been proposed early on as a possible and plausible
explanation for non-repeating FRBs \citep{Falcke2013}. More specifically,
this model suggests that an isolated and magnetised supramassive neutron
star, \ie a neutron star whose mass is the maximum mass for nonrotating
configurations, collapses when it has lost sufficient angular momentum
via the emission of EM energy via dipolar radiation. When this happens,
the rotating star disrupts its magnetosphere and launches a coherent EM
emission in the radio band \citep{Falcke2013}.

We have here explored the validity of this model going beyond the
numerical modelling presented by \citet{Dionysopoulou:2012pp}, who
considered the gravitational collapse of a magnetised but nonrotating
neutron star within a resistive-MHD framework in general relativity. In
particular, we have performed accurate numerical simulations of
collapsing neutron stars adopting a framework similar to that of
\citet{Dionysopoulou:2012pp}, but considering here a large number of
rotating neutron-star models that differ either in rotation rate or in
the initial magnetisation.

Overall, and as observed also in the case of nonrotating stars, we have
found that when rotation is involved the disruption of the magnetosphere
still takes place on a dynamical timescale. The EM emission is
characterised by a precursor signal, followed by a main emission pulse
and then by an exponentially decaying signal, typical of the ringdown of
black holes from EM perturbations. All the different peaks in EM wave
train have a sub-millisecond separation and thus highlight that the
blitzar model can easily accommodated multi-peaked FRB signals such as
the one for FRB \citep{Champion2016}. Furthermore, should the EM emission
from an FRB be accompanied by an exponentially decaying EM signal, it
would provide unambiguous evidence that a black hole has indeed been
produced together with the FRB.

When considering the EM energy properties of the blitzar emission we have
found that this is only very weakly dependent on the initial stellar
rotation rate, at least for the rotation rates considered here that
go up to spin frequency of the fastest known pulsar, \ie PSR J1748-2446ad. 
Similarly, the timescale for the EM emission to have
decreased by four orders of magnitude is of the order of one millisecond,
in reasonable agreement with the observations of FRBs. Exploiting this
property and the results of a number of simulations of stellar models
that differ only in the initial magnetic-field strength we have been able
to show that the radiated EM energy scales quadratically with the
magnetic field and that the collapse is able to release in Poynting flux
about $2\%$ of the EM energy initially stored in the magnetosphere. This
magnetic-energy efficiency is essentially independent of the initial
magnetic field and only very weakly dependent on the rotation rates. This
results has therefore allowed us to derive a phenomenological expression
for the emitted EM energy so that once an FRB of a given energy is
measured, it would in principle be possible to set constraints on either
the radius of the collapsing star or on its magnetic field if the
emission is indeed produced by a blitzar.

Before concluding we should stress that while the simulations reported
here represent a significant progress in the modelling of non-repeating
FRBs as blitzars, they also have a number of limitations that call for
additional studies and improvements. First, the results presented here
are simply ``bolometric'' in the sense that we are simply computing the
EM energy emitted from the collapsing process in terms of the Poynting
flux measured at large distances from the source. No attempt has been
made to go beyond the curvature-radiation model of \citet{Falcke2013} to
discuss how the radiated bulk energy is transformed into the observed
radio emission in FRBs. While this is beyond the scope of this paper, it
is part of our programme of modelling blitzar emission. Second, we have
here considered a simplified equation of state to describe the nuclear
matter and a single value for the mass of the
collapsing star. It would be of great interest to explore how the results
presented here change when the stellar models of different masses and
different radii are considered. 

Finally, while a resistive-MHD approach is a versatile approach to
describe the transition between a highly-conductive neutron-star interior
and the electrovacuum that should characterise pulsars, it still
represents an approximation that can be further improved by varying
the choice for the initial electric field, the prescription for the
conductivity profile, and possibly the match to a force-free
exterior. All of these options will be explored in our future work.

\section*{Acknowledgements}

It is a pleasure to thank Heino Falcke for useful discussions and input
\tbf{and to the anonymous referee for the considerations on the
  dirty-blitzar scenario.} Partial support comes from the ERC Synergy
Grant ``BlackHoleCam'' (Grant 610058), from ``NewCompStar'', COST Action
MP1304, from the LOEWE-Program in HIC for FAIR, from the European Union's
Horizon 2020 Research and Innovation Programme (Grant 671698) (call
FETHPC-1-2014, project ExaHyPE).  AN is supported by an Alexander von
Humboldt Fellowship.  The simulations were performed on SuperMUC at
LRZ-Munich, on LOEWE at CSC-Frankfurt and on Hazelhen at HLRS in
Stuttgart.





\section*{}
\bibliographystyle{mnras}
\bibliography{aeireferences}

\begin{figure*}
  \centering
  \includegraphics[width=0.44\textwidth]{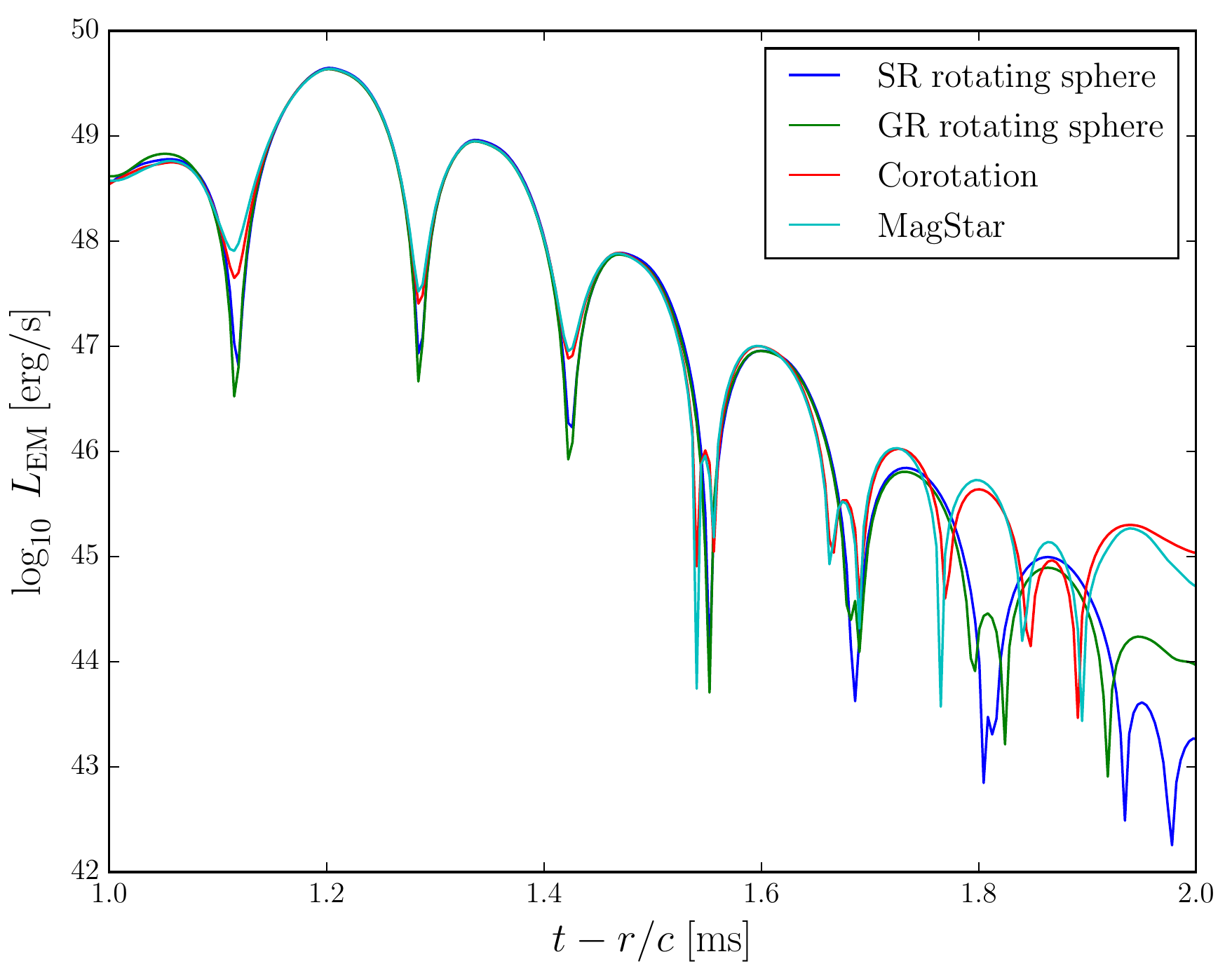}
  \includegraphics[width=0.44\textwidth]{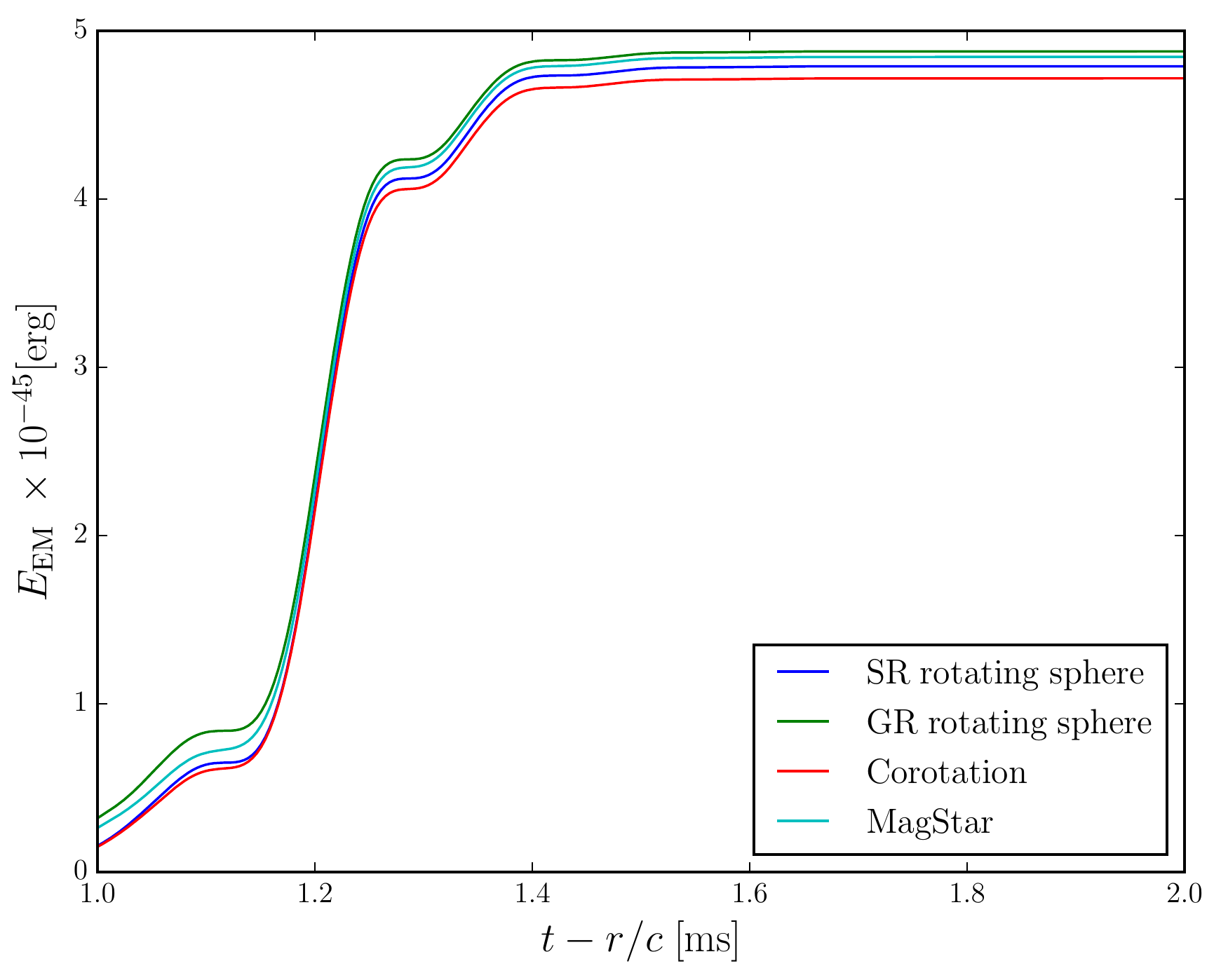}
  \caption{\textit{Left panel:} EM luminosity for the different choices
    of the initial electric field for model \ttt{F800.B13} as
    extracted at $r\simeq 205\,{\rm km}$. Note that the differences among
    the different prescriptions are effectively very small.
    \textit{Right panel:} The same as Fig. \ref{fig:poynting_comparison}
    but for the emitted EM energy. }
  \label{fig:poynting_comparison}
\end{figure*}


\appendix

\section{Initial Electric field}
\label{appen}

In section \ref{sec:nsaid} we discussed the different possible
choices for the initial electric field; in this Appendix we explore the
impact they have on the energetic output from a collapsing reference
model. To maximise such impact, we consider a rather extreme example,
that is, a star rotating at $800\,{\rm Hz}$ and with an initial magnetic
field of $10^{13}\,{\rm G}$, \ie model \ttt{F800.B13}.

We recall that in view of its infinite conductivity, the electric field
for the neutron-star interior is given by the ideal-MHD condition \ie
$E^i = - \epsilon^{ijk} (v_{\rm c})_j B_k$. This interior solution needs
to be matched at the stellar surface and then extended to the outer edge
of the computational domain in such a way that the closed magnetosphere
is corotating and, at the same time, is compatible with our electrovacuum
representation of the stellar exterior. In practice, since any choice of
an electric field would numerically introduce electric charges, the main
goal of the prescription is that of reducing the total net charge and any
spurious effect that may arise at the stellar surface 

The first choice is to employ the analytic description of the electric
field outside a rotating, magnetised and charged sphere in special
relativity proposed by \citet{Ruffini73} (we refer to this solution as to
``SR rotating sphere''). The second choice is the general-relativistic
equivalent of this solution, namely the electric field coming from the
analytical description of a rotating magnetised sphere in general
relativity \citep{Rezzolla2001, Rezzolla2001_err}; this solution is
further modified with the addition of monopolar and quadrupolar terms in
order to account also for the net electric charge of the star
\citep{Ruffini73}. In this way, the modified solution of
\citet{Rezzolla2001} is the one that minimises the exterior charge
density and is the one that was used throughout the paper (we refer to
this solution as to ``SR rotating sphere''). The third and fourth choices
that we have considered are given respectively by a corotation solution
also for the stellar exterior (we refer to this solution as to
``Corotation'') and by the default solution provided by the
\ttt{Magstar} code (we refer to this solution as to ``MagStar'').

The left panel of Fig. \ref{fig:poynting_comparison} offers a comparison
of these four prescriptions for the electric field in terms of the EM
luminosity\footnote{Note that because of its higher rotation rate and
  magnetic field, model \ttt{F800.B13} has a larger luminosity than
  what shown for the other models in
  Fig. \ref{fig:Poynting_comp}.}. Clearly, the main features of the
luminosity produced during the collapse are all very similar, with some
differences becoming visible only in the late stages of the evolution,
when the luminosity has decreased by about five orders of magnitude and
is close to the constant noise level of our code. Similarly, the right
panel of Fig. \ref{fig:poynting_comparison} shows a comparison among the
total emitted EM energies for the different prescriptions. Also in this
case, the different luminosities are very similar and in the range $\sim
4.5 - 5 \times 10^{45}\,{\rm erg}$. Note that the differences reported
are actually smaller than the errors introduced when extracting the
radiation at different coordinate radii.

In view of the results in Fig. \ref{fig:poynting_comparison}, we can
safely conclude that the results of our analysis are robust and not
influenced by the particular choice of the initial electric field.


\label{lastpage}
\end{document}